\documentclass{aa}
\usepackage{dirtytalk}
\usepackage{natbib}
\usepackage{graphicx}
\usepackage{epsfig}
\usepackage{txfonts}
\usepackage{hyperref}
\usepackage{url}
\usepackage{xcolor}
\hypersetup{
colorlinks,
linkcolor={red!80!black},
citecolor={blue!80!black},
urlcolor={blue!80!black}
}

\def\cm3{cm$^{-3}$}
\def\kms{km~s$^{-1}$}

\def\msun{M$_{\odot}$}

\def\one{\ts {\,\sc i}}
\def\two{\ts {\,\sc ii}}

\def\beq{\begin{equation}}
\def\eeq{\end{equation}}

\def\lesssim{\mathrel{\hbox{\rlap{\hbox{\lower4pt\hbox{$\sim$}}}\hbox{$<$}}}}
\def\gtrsim{\mathrel{\hbox{\rlap{\hbox{\lower4pt\hbox{$\sim$}}}\hbox{$>$}}}}

\def\one{{\,\sc i}}
\def\two{{\,\sc ii}}

\def\v1d{{\tt V1D}}

\def\cmfgen{{\tt CMFGEN}}
\def\longpol{{\tt LONG\_POL}}

\newcommand{\iso}[2]{\ensuremath{^{#1}\rm{#2}}}

\def\aj{AJ}
\def\pasp{PASP}

\def\apj{ApJ}
\def\apjs{ApJS}

\def\aap{A\&A}
\def\araa{ARA\&A}

\def\mnras{MNRAS}
\def\nat{Nature}

\def\nifs{\iso{56}Ni}
\def\cofs{\iso{56}Co}

\begin{document}

   \title{Multiepoch VLT\,$-$\,FORS spectropolarimetric observations of supernova 2012aw reveal an asymmetric explosion}
   \titlerunning{Modeling of the SN\,2012aw spectropolarimetric obsvervations}

\author{
   Luc Dessart\inst{\ref{inst1}}
  \and
   Douglas C. Leonard\inst{\ref{inst2}}
  \and
   D. John Hillier\inst{\ref{inst3}}
   \and
   Giuliano Pignata\inst{\ref{inst4},\ref{inst5}}
}

\institute{
Institut d'Astrophysique de Paris, CNRS-Sorbonne Universit\'e, 98 bis boulevard Arago, F-75014 Paris, France.\label{inst1}
  \and
    Department of Astronomy, San Diego State University, San Diego, CA 92182-1221, USA.\label{inst2}
  \and
    Department of Physics and Astronomy \& Pittsburgh Particle Physics, Astrophysics, and Cosmology Center (PITT PACC),  University of Pittsburgh, 3941 O'Hara Street, Pittsburgh, PA 15260, USA.\label{inst3}
\and
Departamento de Ciencias Fisicas - Universidad Andres Bello,  Avda. Republica 252, Santiago, 8320000 Chile.\label{inst4}
\and
Millennium Institute of Astrophysics (MAS), Nuncio Monse\~nor Sotero Sanz 100, Providencia, Santiago, Chile.\label{inst5}
}

   \date{Received; accepted}
  \abstract{We present VLT\,$-$\,FORS spectropolarimetric observations of the type II supernova (SN) 2012aw taken at seven epochs during the photospheric phase, from 16 to 120\,d after explosion. We corrected for interstellar polarization by postulating that the SN polarization is naught near the rest wavelength of the strongest lines -- this is later confirmed by our modeling. SN\,2012aw exhibits intrinsic polarization, with strong variations across lines, and with a magnitude that grows in the 7000\,\AA\ line-free region from 0.1\,\% at 16\,d up to 1.2\,\% at 120\,d. This behavior is qualitatively similar to observations gathered for other type II SNe. A suitable rotation of Stokes vectors places the bulk of the polarization in $q$, suggesting the ejecta of SN\,2012aw is predominantly axisymmetric. Using an upgraded version of our 2D polarized radiative transfer code, we modeled the wavelength- and time-dependent polarization of SN\,2012aw. The key observables may be explained by the presence of a confined region of enhanced \nifs\ at $\sim$\,4000\,\kms, which boosts the electron density in a cone having an opening angle of $\sim$\,50\,deg and an observer's inclination of $\sim$\,70\,deg to the axis of symmetry. With this fixed asymmetry in time, the observed evolution of the SN\,2012aw polarization arises from the evolution of the ejecta optical depth, ionization, and the relative importance of multiple versus single scattering. However, the polarization signatures exhibit numerous degeneracies. Cancellation effects at early times imply that low polarization may even occur for ejecta with a large asymmetry. An axisymmetric ejecta with a latitudinal-dependent explosion energy can also yield similar polarization signatures as asymmetry in the \nifs\ distribution. In spite of these uncertainties, SN\,2012aw provides additional evidence for the generic asymmetry of type II SN ejecta, of which VLT\,$-$\,FORS spectropolarimetric observations are a decisive and exquisite probe.
}

   \maketitle

\section{Introduction}

Photometry and spectroscopy provide critical information on type II supernova (SN). They characterize, for example, the timescale over which the energy stored in the ejecta is released, the color of the escaping radiation, the level of ionization of the spectrum formation region, and the composition of the ejecta. Such data do not, however, provide direct constraints on the geometry of the explosion. Lacking spatially resolved observations for type II-Plateau (II-P) SNe, spectropolarimetry is the best technique to provide constraints on the morphology of SN ejecta, especially at early times \citep{shapiro_sutherland_82}. The polarization is linear, caused by scattering with free electrons, and is typically less than 1\% in the continuum \citep{wang_wheeler_rev_08}. The current spectropolarimetric dataset of type II SNe is limited to 10\,$-$\,20 objects, mostly because of the scarcity of nearby events. Indeed, high signal-to-noise ratio observations are required to reveal such a low-level of polarization.

 SN\,1987A is a hydrogen-rich, core collapse SN classified as type II-peculiar. Unlike type II-P SNe, which arise from red-supergiant (RSG) star explosions, the progenitor of SN\,1987A was a blue supergiant star. It is the first SN for which multiepoch multiband polarimetric observations were obtained \citep{mendez_87A_pol_88,1988SvAL...14..163V,1988MNRAS.234..937B,CBM88_87A_pol,Jef91_obs_pol,WWH02_1987A}. The 0.5\% continuum polarization level observed in the optical during the first two months is understood to arise from scattering with free electrons in a prolate or oblate ejecta \citep{hoeflich_87A_91,jeffery_87_pol_91}. The polarization angle shows variations in time and across lines, suggesting some departures from axial symmetry. Overall, an asymmetric distribution of \nifs\ may also account for the observed polarization \citep{chugai_87A_92}.

Subsequently, spectropolarimetric observations have been gathered for a number of nearby type II SNe, but now of the plateau type, starting with SN\,1999em \citep{leonard_99em_specpol_01}. For SN\,2004dj, \citet{leonard_04dj_06} collected spectropolarimetric observations throughout the photospheric and the nebular phase, up to about a year after explosion. The intrinsic polarization of the SN was revealed unambiguously by the strong variation in polarization across lines. Furthermore, the continuum polarization  evolved from being small during the photospheric phase to being maximum at the onset of the nebular phase, subsequently decreasing as the inverse of the time squared. \citet{chugai_04dj_06} proposed that the polarization was associated with the excess excitation and ionization originating from the presence of high velocity \nifs\ fingers or blobs. Although intimately associated with ejecta asymmetry, \citet{DH11_pol} suggest that the polarization peak at the onset of the nebular phase is caused by a radiative transfer effect as the ejecta turns optically thin, while its subsequent evolution reflects the drop in ejecta optical depth. \citet{chornock_pol_10} presented spectropolarimetric observations for a few more type II SNe, confirming the general trend observed for SN\,2004dj. It thus appears that type II SNe typically show a temporal increase in continuum polarization through the photospheric phase, reaching a maximum at the onset of the nebular phase, and subsequently dropping.

To supplement this dataset, we initiated a VLT\,$-$\,FORS spectropolarimetric program in 2008. Over a number of years, we gathered multiepoch high quality observations for SNe 2008bk, 2012aw, and 2013ej, as well as a more sparse dataset for a few additional objects. Although this dataset has been presented in a  terse format previously \citep{leonard_08bk_12,leonard_iauga_15}, we are now in a position to present the observations in more detail, and with modeling results. We start in this paper with SN\,2012aw.

In the next section, we summarize the results from the polarization modeling of \citet{DH11_pol}. Section~\ref{sect_modeling} presents the numerical approach used for the present study, which is also presented in detail in Hillier \& Dessart (in prep.). As in \citet{DH11_pol}, we adopt a 2D axisymmetric ejectum but we now model the polarized spectrum for the whole optical range rather than focus on isolated lines and the overlapping continuum. In Section~\ref{sect_obs} we present the spectropolarimetric observations of SN\,2012aw that we collected with VLT\,$-$\,FORS while section~\ref{sect_res} presents the results of our modeling of SN\,2012aw covering both the photospheric and the nebular phase. In Section~\ref{sect_degen}, we investigate the degeneracy of the polarization results by investigating the impact of our assumptions (magnitude of the asymmetry, adoption of the mirror symmetry with respect to the equatorial plane, and nature of the asymmetry). We present our conclusions in Section~\ref{sect_conc}.

\section{Summary of results from Dessart \& Hillier (2011)}
\label{sect_dh11_pol}

\citet{DH11_pol} studied the linear polarization in 2D axisymmetric Type II SN ejecta using a long-characteristic code as well as a Monte Carlo code. They explored how the intrinsic continuum and line polarization change with wavelength, bound-bound transition, albedo, SN age, or some properties of the asymmetry.  Below, we summarize their results since they are relevant for the present study and the interpretation of polarization signatures in general.

Even though electron scattering produces a gray opacity, the absorption opacity varies between photoionization edges and causes a variation of the albedo with wavelength. Consequently, the continuum polarization in type II SNe may vary across the optical range. In the modeling, we take this effect into account by separating the different sources of opacity at each wavelength, namely electron scattering on the one hand and on the other hand the bound-bound and bound-free processes (see also \citealt{hillier_94,hillier_96}). Scattering in lines (i.e., bound-bound transitions), and especially resonance lines, is assumed to be unpolarized. In reality, line scattering has a polarizing effect but it is much weaker than the polarizing effect caused by free electrons \citep{jeffery_87_pol_91}.

The residual polarized flux arises from a non-cancellation of the local polarization integrated on the plane of the sky. A net polarization can result from an asymmetric distribution of scatterers, but it can also stem from an asymmetric distribution of the flux (for example due to optical-depth effects; \citealt{hillier_94}; \citealt{DH11_pol}; \citealt{vlasis_2n_16};
\citealt{2015MNRAS.450..967B}).

Even for large asphericity, a large polarization may not result because of strong cancellation effects. This situation holds particularly at early times because of the confinement of the spectral formation region, which favors a low residual polarization. The integrated polarized flux is also inhibited by multiple scatterings since these tend to randomize the scattering directions, making the radiation field more isotropic. These effects suggest that SN ejecta should produce a small (or at least a smaller) continuum polarization at earlier times, whatever the level of asymmetry. At nebular times, when the ejecta optical depth is below unity along all sight lines, photons typically scatter once with free electrons. Because of the weaker cancellation effects, this situation produces a greater residual polarization for a given asymmetric configuration. Under optically thin conditions, the continuum polarization scales linearly with continuum optical depth \citep{Brown_McLean_77}. Despite these advantages,  the faintness of SNe at nebular times presents a major challenge for spectropolarimetric observations \citep{leonard_04dj_06}.

The effects of optical depth can lead to complex polarization signatures, distinct from what is obtained under optically thin conditions. One such feature is a sign reversal of the polarization (equivalent to a 90\,deg change in the polarization angle) in the continuum across the optical range (in part driven by the change in albedo between the Balmer edge and the Paschen edge), as well as sign reversals across line profiles and the adjacent continuum. In observations, sign reversals (especially if they are weak) may be hard to diagnose if the interstellar polarization is not known accurately. Sign reversals have been seen in spectropolarimetric observations of some type II-P SNe \citep{chornock_pol_10}.

In optically thick lines such as H$\alpha$, the polarization is generally zero somewhere between the location of maximum absorption in the P-Cygni trough and the location of maximum flux near line center (see for example Figs.~8 and 16 in \citealt{DH11_pol}, in cases where the zero crossing is associated with a sign reversal of the polarization). This location may thus be useful for constraining the interstellar polarization, especially if other lines can also be used for that purpose (for example the Ca\,\two\ near-infrared triplet or H$\beta$). In contrast, weak lines may not cause much reduction to the continuum polarization (for example H$\delta$) and thus only induce an inflection in the continuum polarization level. In the opposite regime of strong line overlap, for example in regions of line blanketing, both the total flux and the polarized flux are strongly reduced.

For a fixed ejecta asymmetry, the continuum polarization is expected to reach a maximum when the core is revealed. This is understood as stemming from the progressive growth of the extent of the spectrum formation region during the photospheric phase and the drop in envelope optical depth (favoring single instead of multiple scatterings). It may occur at fixed asymmetry, if the asymmetry increases toward the inner ejecta regions, and may even occur if the asymmetry is confined to the outer, H-rich ejecta layers. Such a strong increase in continuum polarization has been seen in a few type II SNe \citep{leonard_04dj_06,chornock_pol_10}, and this feature is also present in the SN that we discuss in this paper. At nebular times, when the total electron scattering optical depth of the ejecta is below unity, the continuum polarization is expected to follow a $1/t^2$ evolution for a fixed ionization (since the asymmetry and viewing angle do not change with time; see also \citealt{Brown_McLean_77}), following the drop in ejecta optical depth. In SN ejecta, however, deviations from this could arise if the asymmetry varies with depth and if the ionization evolves in a complicated fashion -- for example following the growth of the $\gamma$-ray photon mean free path with time.

All of these conclusions apply to the new models presented here using the upgraded long-characteristic code (Section~\ref{sect_modeling_pol}; see also Hillier \& Dessart, in prep.). However, they are now more directly visible since the upgraded code delivers the full polarized spectrum (across the optical or near-infrared range) rather than just the polarization for a single isolated line and its adjacent continuum, as previously done in \citet{DH11_pol}.

\begin{figure}[t]
\epsfig{file=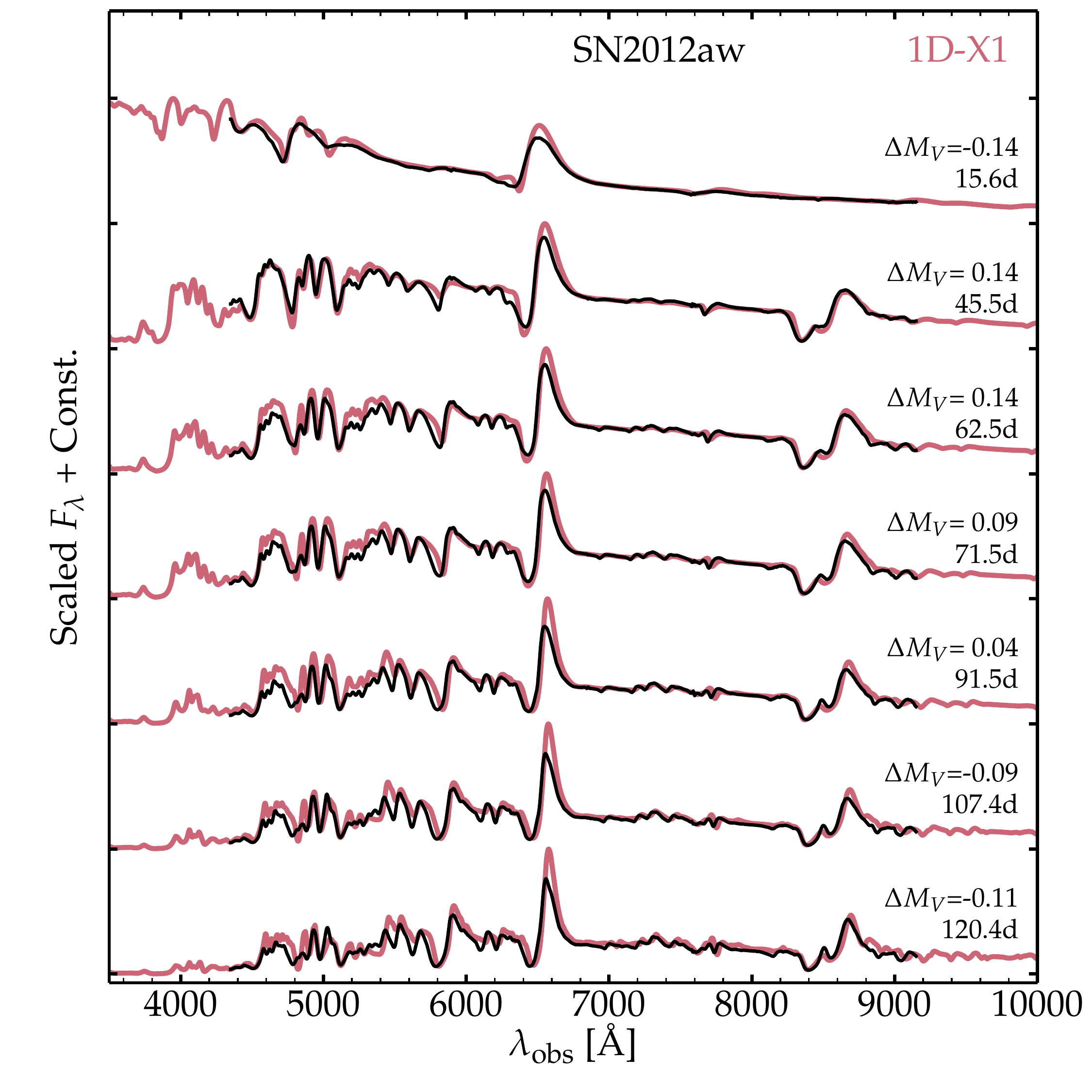,width=9cm}
\caption{Multiepoch spectral comparison between SN\,2012aw and our reference 1D \cmfgen\ model 1D-X1 (see Section~\ref{sect_ref_1d_cmfgen}). The data correspond to the observed total flux for SN\,2012aw obtained with VLT\,$-$\,FORS in spectropolarimetric mode. The model is redshifted, reddened, and then normalized to the observed flux at 7000\,\AA. The $V$-band brightness offset, calculated from the observed photometry (\citealt{bose_12aw_13}; corrected for distance and extinction) and the model photometry, is given for each epoch and is typically on the order of 0.1\,mag. This model is named x1p5 in \citet{HD19} and is our best match model for SN\,2012aw.
\label{fig_spec_montage_1dx1}
}
\end{figure}
\begin{figure}[t]
\epsfig{file=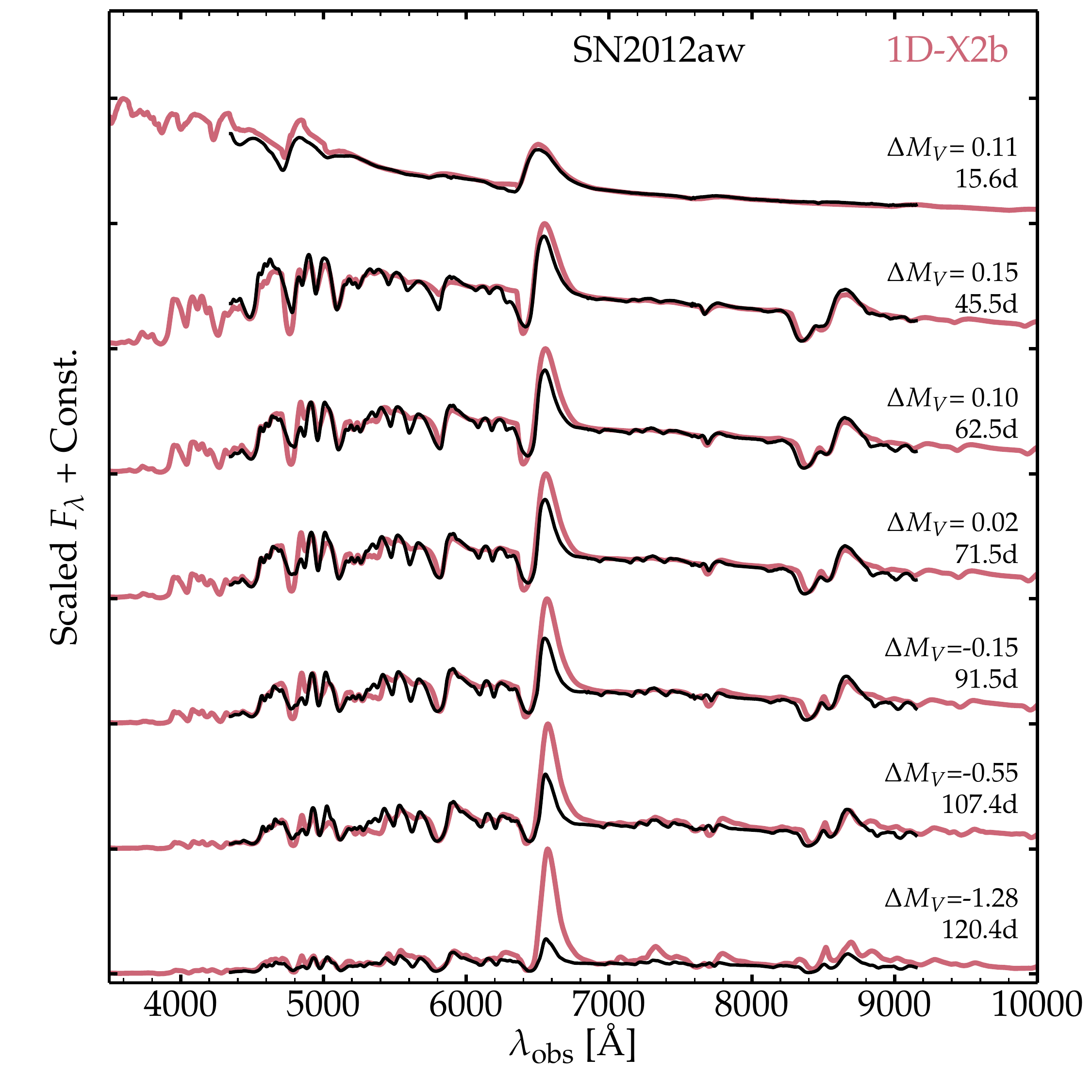,width=9cm}
\caption{Same as for Fig.~\ref{fig_spec_montage_1dx1}, but now showing the results for model 1D-X2b. Compared to model 1D-X1, model 1D-X2b is designed to be more strongly mixed, and to possess a \nifs\ shell around 4000\,\kms. With these modifications it deviates more strongly from the observations of  SN2012aw.
\label{fig_spec_montage_1dx2b}
}
\end{figure}

\begin{figure}[h!]
\epsfig{file=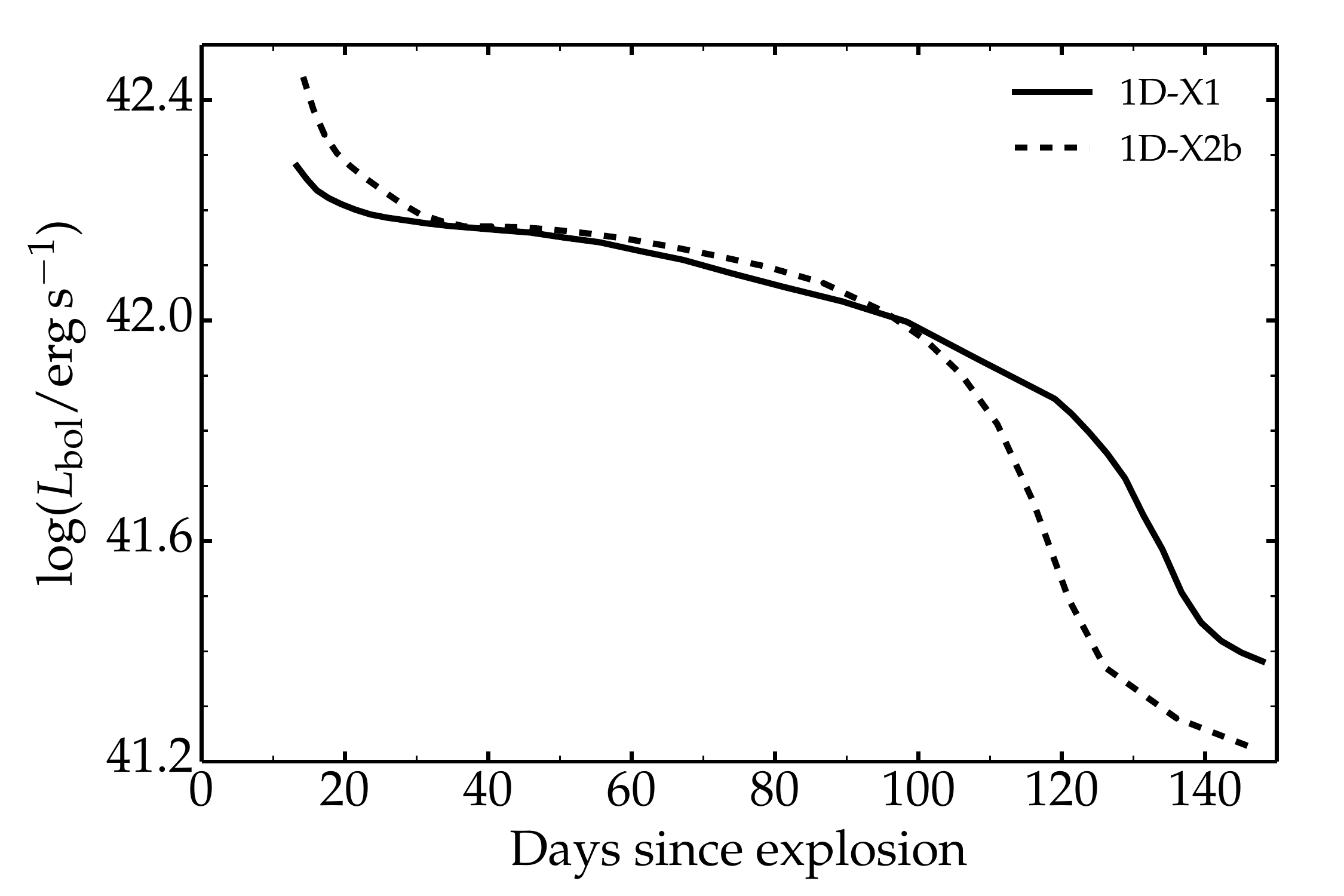, width=8cm}
\epsfig{file=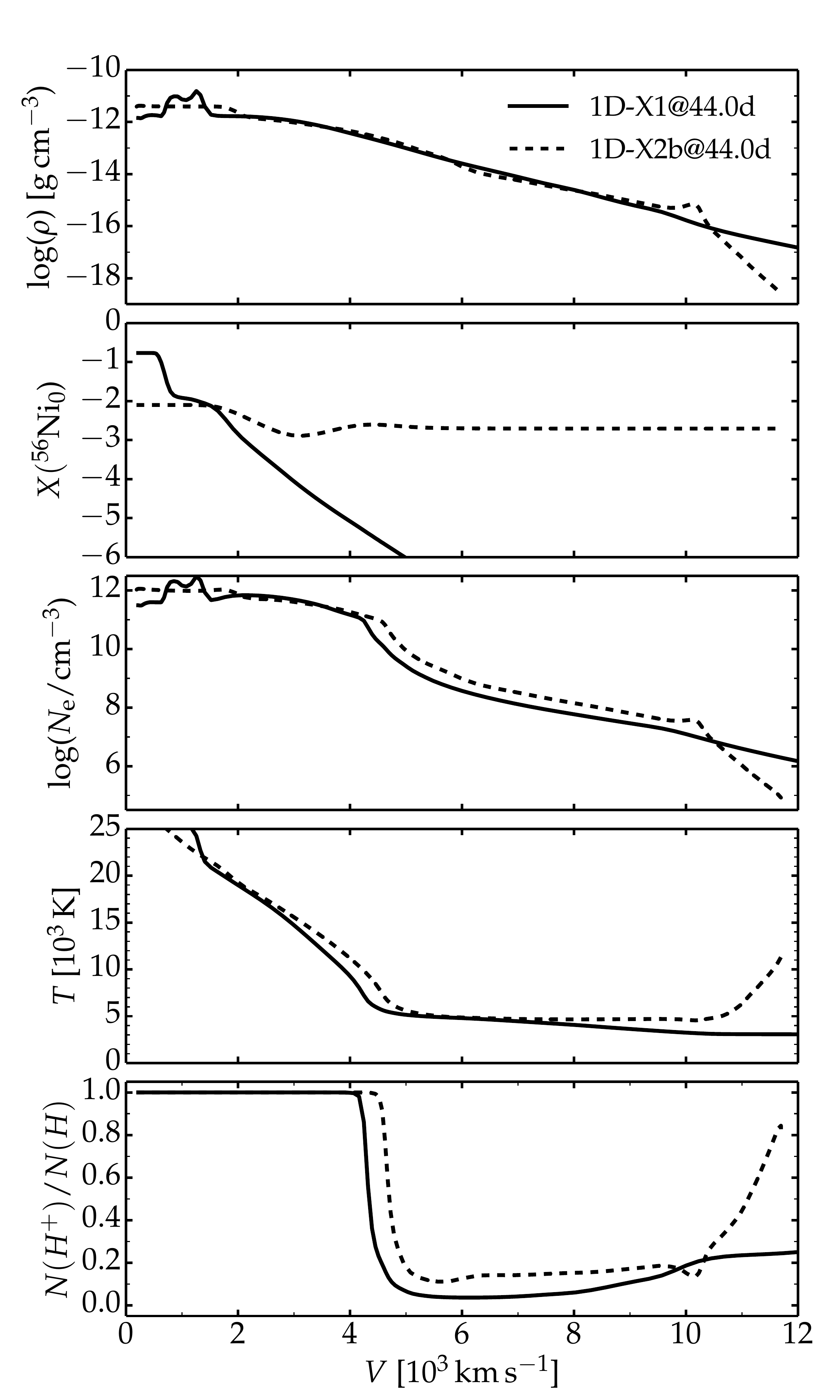, width=8cm}
\caption{Top: Bolometric light curve for the 1D \cmfgen\ models 1D-X1 and 1D-X2b. Bottom: Mass density, initial \nifs\ mass fraction, electron density, gas temperature, and hydrogen ionization fraction vs. velocity for models 1D-X1 (solid) and 1D-X2b (dashed) at 44\,d after explosion. These two models vary in the outer ejecta structure, the inner core structure, and the \nifs\ mixing adopted (strong or weak, presence or absence of a \nifs-rich shell at a Lagrangian mass of 12\,\msun, corresponding to a velocity of about 4000\,\kms\ in the resulting ejecta). The two models have the same mass-density profile outside of the inner ejecta (with the exception of the regions beyond 10000\,\kms), but different electron-density profiles because of the different \nifs\ profiles.
\label{fig_1d_prop}
}
\end{figure}

\section{Modeling approach}
\label{sect_modeling}

The spectropolarimetric modeling we perform in this study requires a few preparatory steps. All simulations are based on 1D nonlocal thermodynamic equilibrium (nonLTE) time-dependent radiative transfer simulations of RSG star explosion models performed with \cmfgen, computed recently for a separate study on the diversity of Type II SNe \citep{HD19}. These \cmfgen\ simulations are used as initial conditions for the 2D polarized radiative transfer modeling. We use an updated version of the long-characteristic code described in \citet{DH11_pol}. The main improvements are the computation of the full optical polarized flux as well as extra flexibility for setting up the 2D axisymmetric ejecta (see Hillier \& Dessart, in prep.).

Here, a 2D axially-symmetric ejecta is built using various approaches. The first approach is to introduce a latitudinal scaling to the density (i.e., mass density, ion density, atom density, and electron density), opacity, and emissivity computed from a 1D \cmfgen\ model. The second approach is to \say{stretch} these quantities in radial space, by an amount that depends on latitude. One limitation of these two approaches is that the adopted scaling or radial offset may not reflect well the properties of axisymmetric and asymmetric ejecta. In the third approach, we assign distinct 1D \cmfgen\ models to a range of latitudes so all latitudes are characterized by a physical, albeit 1D, model. One may use two models (defining the properties of the 2D ejecta along the equator and the pole) or a larger set of models to introduce smaller scale asymmetries covering a smaller opening angle.

In this work, we used a combination of approaches. For the first epoch of spectropolarimetric observations of SN\,2012aw, we explored the influence of a latitudinal scaling or a latitudinal radial displacement of a 1D model (options 1 and 2 above). This somewhat artificial approach is not followed further. Instead, the bulk of the modeling is based on a mapping in latitude of two distinct 1D \cmfgen\ models to produce a large scale asymmetry of the ejecta. This approach is also applied from early to late times and thus allows one to see the evolution of the SN polarization for a given asymmetric ejecta. Below, we review the properties of the 1D \cmfgen\ models and the various 2D asymmetric (but axisymmetric) ejecta that we build from these, before describing in detail how the polarization calculations are performed. The nomenclature as well as the distinguishing properties of the 1D and 2D simulations performed in this study are summarized in Table~\ref{tab_set}.

\begin{table*}
\caption{Summary of 1D and 2D model properties. The first column gives the dimensionality of the simulation, 1D referring to the spherically-symmetric nonLTE time-dependent simulations with \cmfgen\ and 2D to the polarized axially-symmetric steady-state radiative transfer simulations with \longpol. The second column gives the model name. The third column provides some characteristics of each simulation. The rightmost column indicates in which section the corresponding simulation is discussed. All 2D simulations adopt mirror symmetry with respect to the equatorial plane unless otherwise stated.
\label{tab_set}
}
\begin{center}
\def\arraystretch{1.4}
\begin{tabular}{lllr}
\hline
Dim.       &  Model name  & Characteristics   & Section \\
\hline
\hline
\multicolumn{4}{c}{\cmfgen\ simulations} \\
\hline
1D  & 1D-X1  & $E_{\rm kin}=1.2 \times 10^{51}$\,erg;  $M_{\rm ej}=12.1$\,\msun;  $M(^{56}$Ni$)=$\,0.056\,\msun.    &   \ref{sect_ref_1d_cmfgen} \\
\hline
1D  & 1D-X2b  & $E_{\rm kin}=1.3 \times 10^{51}$\,erg;  $M_{\rm ej}=12.4$\,\msun;  $M(^{56}$Ni$)=$\,0.047\,\msun. &   \ref{sect_ref_1d_cmfgen} \\
       &                 &  Presence of a \nifs-rich shell at 4000\,\kms\     & \\
\hline
1D  & 1D-X2B  & $E_{\rm kin}=1.9 \times 10^{51}$\,erg;  $M_{\rm ej}=12.4$\,\msun;  $M(^{56}$Ni$)=$\,0.054\,\msun. &   \ref{sect_expl} \\
\hline
\hline
\multicolumn{4}{c}{\longpol\ simulations} \\
\hline
2D  & 2D-X1-SCL0p5  & Model X1 with a latitudinal density scaling; $A=$\,0.5  & \ref{sect_prep_lat_scl}\,$-$\,\ref{sect_scl}   \\
2D  & 2D-X1-SCL4p0  & Model X1 with a latitudinal density scaling; $A=$\,4.0  & \ref{sect_prep_lat_scl}\,$-$\,\ref{sect_scl}   \\
\hline
2D  & 2D-X1-STR0p25  & Model X1 with a latitudinal radial stretching; $A=$\,0.25 & \ref{sect_prep_lat_str}\,$-$\,\ref{sect_str}\\
2D  & 2D-X1-STR0p5   &  Model X1 with a latitudinal radial stretching; $A=$\,0.5   & \ref{sect_prep_lat_str}\,$-$\,\ref{sect_str}\\
\hline
2D  & 2D-X1-X2b  & Combination of models 1D-X1 and 1D-X2b    & \ref{sect_prep_lat_multi}\,$-$\,\ref{sect_multi}\\
       &                       & \hspace{0.5cm} \indent 1D-X2b covers polar angles 0 to $\beta_{1/2}=$\,22.5\,deg.    & \\
      &                       & \hspace{0.5cm} 1D-X1 covers polar angles 22.5 to 90\,deg.   & \\
      &                        &  Influence of variations in opening angle $\beta_{1/2}$ & \ref{sect_cone} \\
      &                        &  Influence of assumption of top/bottom symmetry & \ref{sect_mirror} \\
\hline
2D  & 2D-X1-X2B  & Combination of models 1D-X1 and 1D-X2B    & \ref{sect_expl}\\
       &                       & \hspace{0.5cm} 1D-X2B covers polar angles 0 to $\beta_{1/2}=$\,22.5\,deg.    & \\
      &                       & \hspace{0.5cm}  1D-X1 covers polar angles 22.5 to 90\,deg.   & \\
\hline
\end{tabular}
\end{center}
\end{table*}

\begin{figure*}
\begin{center}
\epsfig{file=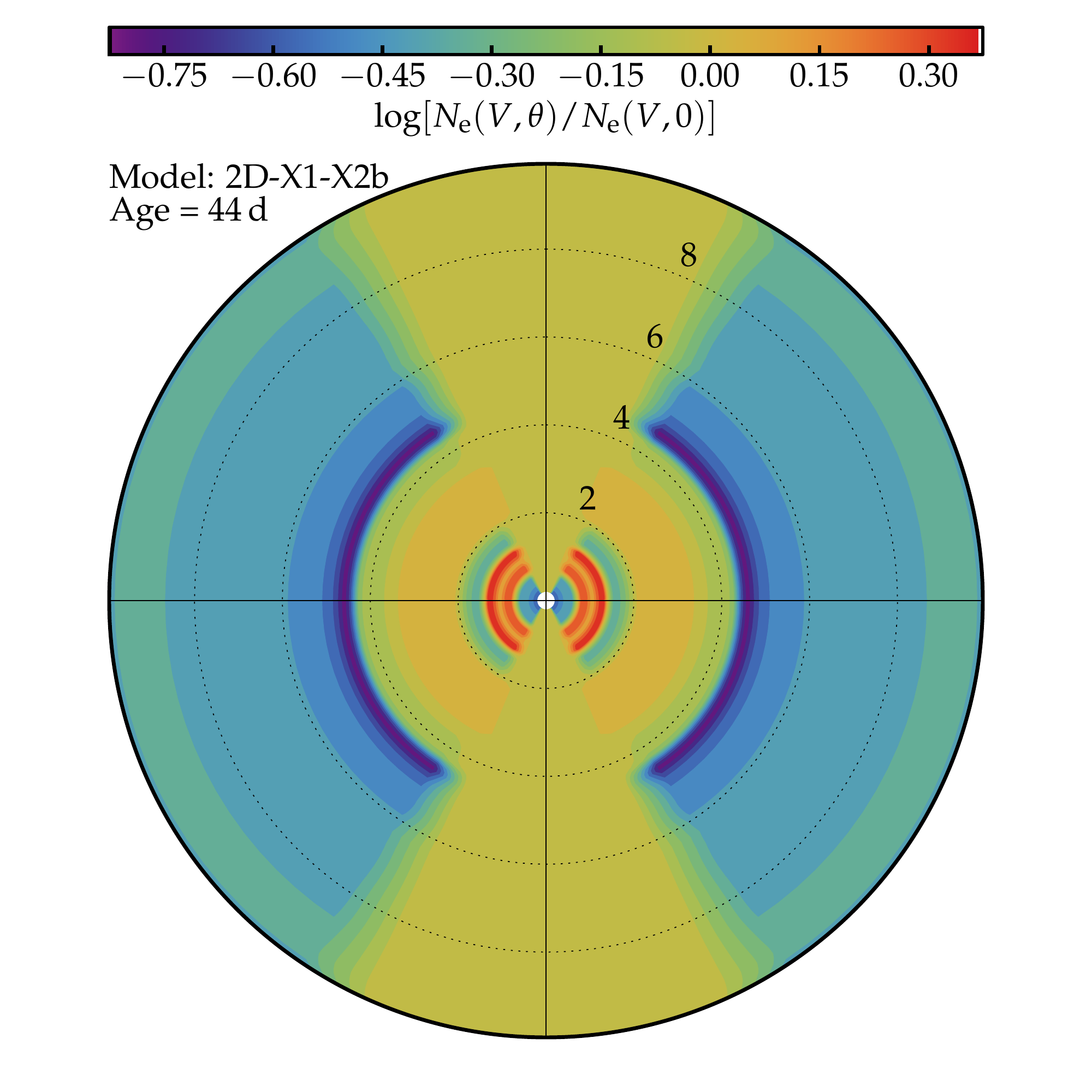, width=8cm}
\epsfig{file=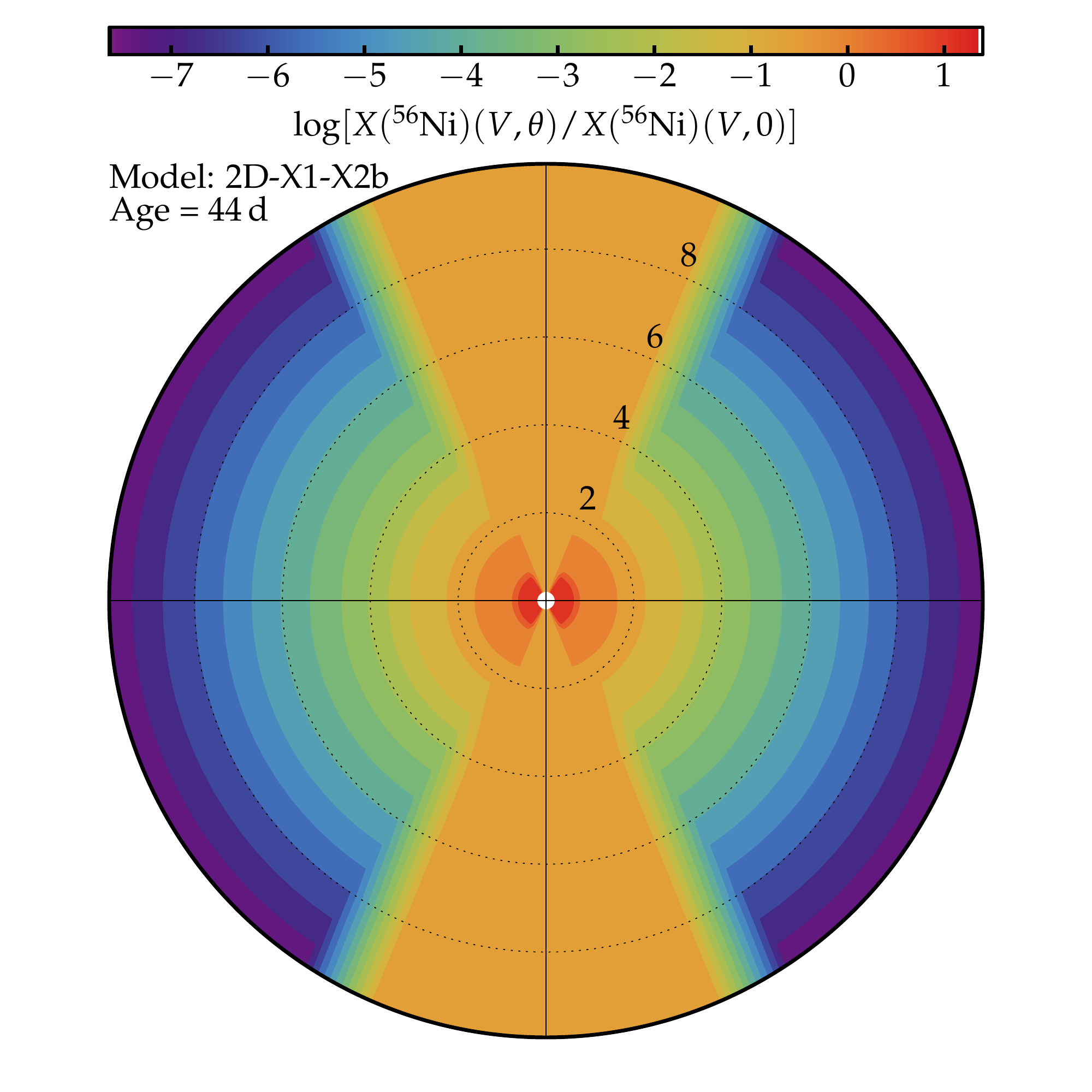, width=8cm}
\end{center}
\vspace{-0.5cm}
\caption{2D distribution of the free-electron density $N_{\rm e}$ (left) and of the \nifs\ mass fraction (right; the \nifs\ mass fraction is in practice always lower than about 0.01 beyond 1000\,\kms\ along all latitudes -- see Fig.~\ref{fig_1d_prop}), normalized to their corresponding values along the pole, for model 2D-X1-X2b at 44\,d after explosion.  We use model 1D-X2b up to $25-30$\,deg, and model 1D-X1 beyond. The vertical corresponds to the axis of symmetry, so polar angle $\beta$ of zero and 180\,deg, and the horizontal corresponds to the equatorial direction.  This axisymmetric ejecta has mirror symmetry with respect to the equatorial plane. The black label along the one o'clock direction gives the radial velocity in units of 1000\,\kms. The ejecta asymmetry stems from the asymmetric distribution of \nifs, as well as differences in the outer ejecta (through the presence of a dense RSG progenitor atmosphere or not) and differences in the inner ejecta (complete mixing and smoothing of the He core material) -- see Fig.~\ref{fig_1d_prop} and Section~\ref{sect_prep_lat_multi} for details.
\label{fig_init_2d_x1p5b3}
}
\end{figure*}

\subsection{The reference 1D \cmfgen\ model}
\label{sect_ref_1d_cmfgen}

SN\,2012aw was studied by \citet{HD19} as part of a large investigation on the origin of the diversity of type II SNe, in particular in terms of $V$-band decline rate and spectral diversity during the photospheric phase. The reference model we use for SN\,2012aw is named here 1D-X1 (originally called x1p5 in \citealt{HD19}) and corresponds to a 15\,\msun\ zero age main sequence star,  exploded to yield an ejecta kinetic energy of $1.2 \times 10^{51}$\,erg, an ejecta mass of 12.12\,\msun, and a \nifs\ mass of 0.056\,\msun. Figure~\ref{fig_spec_montage_1dx1} illustrates the agreement in the optical range (the data are from our spectropolarimetric VLT\,$-$\,FORS program) between model 1D-X1 and SN\,2012aw, and also the fact that the model is within about 0.1\,mag of the observed $V$ band magnitude throughout the high brightness phase (there is a similar offset at nebular times; see \citealt{HD19}). This agreement suggests that a spherical ejecta model yields a satisfactory match to the photometric and spectroscopic observations of SN\,2012aw. The deviations from spherical symmetry that we introduce must therefore remain small enough not to deteriorate the agreement between observations and the results for model 1D-X1.

\subsection{2D-axisymmetric ejecta models}
\label{sect_prep_2d_ejecta}

\subsubsection{Application of a latitudinal scaling to the reference model}
\label{sect_prep_lat_scl}

In this approach, the density $\rho$ at an ejecta location $(r,\beta)$, where $r$ is the local radius and $\beta$ is the polar angle (i.e., the angular separation from the axis of symmetry), is given by
\begin{equation}
\rho(r,\beta) = \rho_{\rm 1d}(r) B (1 + A \cos^2\beta)  \, ,  \label{eq_rho_lat_scaling}
\end{equation}
with $B$ set (in all cases) to $1/(1+A/3)$ (chosen  so that  $\int_0^1 B (1 + A \cos^2\beta) \sin \beta \,d\beta=1$).
With this choice, prolate (oblate) density configurations correspond to a positive (negative but greater than $-1$) $A$ value. Other choices are possible. For example, by varying the exponent on the cosine one can modulate the latitudinal confinement of the asymmetry. This option is not explored here. With this adopted density scaling, and at a given radius $r$, opacities and emissivities scale with $B^2(1 + A \cos^2\beta)^2$, while the electron-scattering opacity scales with $B(1 + A \cos^2\beta)$.

\subsubsection{Application of a radial scaling to the reference model}
\label{sect_prep_lat_str}

The second option is to apply a radial scaling of the 1D \cmfgen\ model and use the same density, opacity, and emissivities as the 1D \cmfgen\ model on that scaled grid. For example, the density (as well as opacities and emissivities) at $r$ along $\beta$ corresponds to its counterpart in the original 1D \cmfgen\ model at $r/B(1+A\cos^2\beta)$. If the density declines outward, positive (negative) $A$ values correspond to prolate (oblate) density configurations. For an ejecta characterized by a power-law density with exponent $n_\rho$, the radial scaling causes a maximum density contrast between pole and equator of $(1 + A)^{n_\rho}$. For $n_\rho=12$ (which is representative of the outer layers of the ejecta in model 1D-X1), this maximum density contrast corresponds to 14.6 for $A=0.25$ and to 130.0 for $A=0.5$ (these values are used in two models presented in Section~\ref{sect_str}). In ejecta regions of constant density, a radial scaling yields no latitudinal variation in density (in realistic ejecta, the density gradient varies with depth so a stretching may correspond to complicated variations in density with both inclination and depth). In contrast, the latitudinal scaling (see Eq.~\ref{eq_rho_lat_scaling}) always introduces a variation in density with latitude, and this variation between pole and equator has the same magnitude at any given depth.

\subsubsection{Combination of the reference 1D \cmfgen\ model with another model}
\label{sect_prep_lat_multi}

In this approach, we build the axisymmetric ejecta assigning distinct 1D \cmfgen\ models to the equatorial and to the polar directions, and interpolate between the models at intermediate latitudes. The mapping of a given model over a range of latitudes is flexible. Multiple models could also be used but here we use only two to facilitate the interpretation. We define $\beta_{1/2}$ as the maximum polar angle over which the ``polar'' model applies.

Hence, besides the reference model 1D-X1, we used an alternate model 1D-X2b. This model deviates from model 1D-X1 both for the progenitor and for the ejecta. The core material within the inner 5\,\msun\ is made completely homogeneous (this implies strong mixing) and is also given a fixed density (conserving total mass and yields), thereby erasing the jump in density present in model 1D-X1 at the He-core edge. The progenitor is enshrouded by a dense extended atmosphere prior to explosion, which produces a lower maximum ejecta velocity as well as  the presence of a dense shell in the outer ejecta. The most important feature is however the presence of a \nifs\ rich shell (in 1D), which we add ``by hand'' at a Lagrangian mass of 12\,\msun, so that it is located around 4000\,\kms\ in the resulting ejecta once homologous expansion is established. In practice, this is done in the radiation-hydrodynamics calculation with \v1d\ (see \citealt{livne_93} and \citealt{HD19} for discussion), at 1000\,s after the explosion trigger (and therefore in mass space since the velocity profile has not settled at that time). Because there is very little mass between 12\,\msun\ (or the ejecta location where the velocity is about 4000\,\kms) and the outermost ejecta layers, an excess in \nifs\ (relative to model 1D-X1) is present throughout the outer ejecta. However, the \nifs\ mass fraction remains very small, typically below 0.01, and the \nifs\ is mixed beyond about 2000\,\kms\ with a material rich in H and He (the material that used to be in the H-rich envelope of the progenitor star). Model 1D-X2b corresponds to a (spherically-symmetric) ejecta with a kinetic energy of $1.29 \times 10^{51}$\,erg, a total mass of 12.43\,\msun, and a \nifs\ mass of 0.047\,\msun.

Figure~\ref{fig_spec_montage_1dx2b} compares the spectral properties of model 1D-X2b with the observations of SN\,2012aw in the optical range. Because of the adjustments made, the resulting SN radiation differs sizably from that obtained for model 1D-X1, and thus deviates from SN\,2012aw. This was the goal, and is in line with the earlier findings of \citet{chugai_04dj_06}. In model 1D-X2b, the stronger mixing and the presence of a \nifs\ shell at 4000\,\kms\ (added to give a source of asymmetry in a 2D ejecta; see below) both lead to a stronger and broader H$\alpha$ line in the second half of the plateau phase. The H$\alpha$ mismatch suggests that there is no strong \nifs\ mixing along all ejecta-centered directions, but that strong \nifs\ mixing, if present, must be limited to a modest solid angle. The lower \nifs\ mass also causes an underestimate of the optical brightness compared to SN\,2012aw. As long as model 1D-X2b covers a small fraction of the whole 2D ejecta volume, the global ejecta and radiation properties should primarily reflect those of model 1D-X1. We compare the ejecta properties and the bolometric light curves of the 1D models 1D-X1 and 1D-X2b in Fig.~\ref{fig_1d_prop}.

In most of the simulations we adopt mirror symmetry with respect to the equatorial plane. Model 1D-X2b is assigned the polar directions and model 1D-X1 is assigned the equatorial direction. The properties at intermediate latitudes is flexible. In the reference setup described in Section~\ref{sect_res}, model 1D-X2b represents the ejecta for all polar angles between zero and 22.5\,deg. Beyond 33.75\,deg, we use model 1D-X1 instead. In between these two angles, we linearly interpolate in $\cos \beta$. This interpolation applies to all relevant quantities for the 2D radiative transfer and in particular the opacities and emissivities. We thus interpolate between two physically consistent 1D \cmfgen\ models (also derived from physically consistent radiation-hydrodynamics simulations of the explosion). This is superior to using one 1D \cmfgen\ model and applying some density scaling or radial scaling on its properties (see previous two sections). Figure~\ref{fig_init_2d_x1p5b3} illustrates some of the properties of the resulting 2D ejecta. The left panel shows the normalized electron density (in the log base 10) at 44\,d after explosion. With this setup, we can explore the configuration for an enhanced \nifs\  abundance confined to a narrow range of latitudes. Such a configuration is known to yield a residual polarization \citep{chugai_04dj_06}.

This setup is not an exact representation of what may occur in nature. Detailed 3D simulations of neutrino-driven explosions suggest that the \nifs\ material may be asymmetrically distributed in a complicated structure of blobs, shells, fingers, elongated toward large velocities, and made of essentially pure \nifs\ \citep{wongwathanarat_15_3d}. This material is macroscopically mixed in space during the dynamical phase of the explosion, so that it may coexist at a given velocity with H-rich material but it is not microscopically mixed with it. In our present approach, we instead apply a macroscopic and a microscopic mixing so that both \nifs\ and H-rich material coexist at a given velocity. This approach is perhaps not as problematic as it seems since, with respect to the polarization of light, it is the influence of \nifs\ on the electron density that matters. As time passes, the $\gamma$-rays associated with the radioactive decay of \nifs\ and \cofs\ are less efficiently trapped locally, and the more so for the \nifs\ advected out to large velocities during the explosion. Thus, the volume influenced by this decay heating extends beyond that occupied by the \nifs-rich material. An illustration of this effect is presented in \citet{d12_snibc}, in particular their Figs.~1 and 2. Hence, the configuration corresponding to an extended cocoon of enhanced ionization (and thus electron density) around any pure \nifs\ blob is well captured by our setup.

\begin{table*}
\caption{Journal of spectropolarimetric observations of SN\,2012aw.
\label{tab_obs}
}
\begin{center}
\begin{tabular}{l@{\hspace{4mm}}
c@{\hspace{4mm}}c@{\hspace{4mm}}c@{\hspace{4mm}}
c@{\hspace{4mm}}c@{\hspace{4mm}}c@{\hspace{4mm}}
c@{\hspace{4mm}}c@{\hspace{4mm}}c@{\hspace{4mm}}
}
\hline
Epoch  & Day$^a$       & UT Date$^b$          &  MJD$^c$ &    Exposure$^d$ & Air Mass$^e$ & Seeing$^f$    \\
\hline
1 & 16.110  & 2012-04-01.200 & 56,018.200 & 1600 (4)  & 1.39\,$-$\,1.55 & 1.2 \\
2 & 45.954  & 2012-05-01.044 & 56,048.044 & 800  (1)  & 1.24\,$-$\,1.25 & 1.2 \\
3 & 62.987  & 2012-05-18.077 & 56,065.077 & 6955 (13) & 1.27\,$-$\,2.17 & 1.0 \\
4 & 71.945  & 2012-05-27.037 & 56,074.037 & 7200 (9)  & 1.26\,$-$\,1.80 & 1.0 \\
5 & 91.945  & 2012-06-16.035 & 56,094.035 & 4860 (9)  & 1.33\,$-$\,1.49 & 1.2 \\
6 & 107.913 & 2012-07-02.003 & 56,110.003 & 3200 (8)  & 1.65\,$-$\,2.76 & 1.0 \\
7 & 120.009 & 2012-07-14.988 & 56,122.988 & 1200 (3)  & 2.25\,$-$\,2.85 & 2.5 \\
\hline
\end{tabular}
\end{center}
Note: All observations were obtained at the VLT using the 300 line/mm grism (``GRIS\_300V'') along with the ``GG435'' order-sorting filter to prevent second-order contamination.  This resulted in a useable spectral range of 4350\,\AA\,$-$\,9200\,\AA.  All observations were made through a 1\farcs0 slit, which delivered a resolution of $\sim$\,12\,\AA, as derived from the FWHM of night-sky lines.  The flux standard used for all observations was BD+26$^\circ$2606 \citep{oke_gunn_83}, observed on the same nights as the science observations.  All observations were made with the position angle of the spectrograph slit set to 0\,deg, which often differed significantly from the parallactic angle \citep{filippenko_82}; when coupled with the relatively narrow slit used, the potential for differential light loss -- and, hence, inaccurate relative flux calibration -- is significant. \\
$^a$ Day since estimated explosion date of 2012-03-16.09\,UT = MJD\,56,002.09 = JD\,2,456,002.09; \citealt{bose_12aw_13}).  \\
$^b$ yyyy-mm-dd, calculated as the midpoint of all of the individual exposures. \\
$^c$ Modified Julian Date (Julian Date - 2400000.5), at the midpoint of all of the individual exposures. \\
$^d$ Total exposure time in seconds, with the number of complete sets of four waveplate positions obtained shown in parenthesis. \\
$^e$ Air mass range for each set of observations. \\
$^f$ Average value of the FWHM of the spatial profile for each epoch, rounded to the nearest 0\farcs1. \\
\end{table*}

\subsection{Our approach for polarization modeling}
\label{sect_modeling_pol}

   The simulations presented in this study were performed with the long characteristic code \longpol\ first implemented to treat continuum \citep{hillier_94} and line \citep{hillier_96} polarization  in the context of multiscattering axisymmetric envelopes of hot star winds. This code was subsequently extended to treat the case of Type II SN ejecta  \citep{DH11_pol} but limited to solving for the polarized flux for an individual line and its overlapping continuum.

   The code has since been improved in two important ways (Hillier \& Dessart 2020, in prep.). First, the code now computes the full optical polarized spectrum and is thus no longer limited to a single line. This implies that line overlap is automatically treated, irrespective of the number of overlapping lines. Second, the code is no longer limited to 2D ejecta produced by the latitudinal density scaling or radial stretching of a precomputed 1D \cmfgen\ model. We can now also combine a set of precomputed 1D \cmfgen\ models, assigning a given model to a specific range of latitudes.  The benefit from this latter approach is the improved physical consistency since the opacities and emissivities imported into \longpol\ are those from the precomputed nonLTE 1D \cmfgen\ models. This is less artificial than prescribing an asymmetry through parametrized scalings. A remaining weakness is that we still interpolate between 1D models at a given latitude. Further, the opacities and emissivities (with the exception of the electron-scattering emissivity) assigned to this 2D ejecta are not computed from a fully consistent 2D nonLTE model. Practically, the 2D radiative transfer is performed as a formal solution (i.e., with the opacities and emissivities fixed). The other benefit is that we can combine any variety of models, differing in explosion energy, composition such as \nifs\ mass or \nifs\ mixing, or clumping.

In \longpol, the polarization is produced by electron scattering and described by the Stokes parameters $I$, $Q$, $U$, and $V$ \citep{chandra_60}. The polarization is thus linear and $V$ is identically zero. We use the same nomenclature as in \citet{DH11_pol} and define the corresponding flux-like polarization quantities $F_I$, $F_Q$, and $F_U$ output by the code (the calculations are done in $I_l$, $I_r$, $U$ and $V$ space where $I_l=0.5(I+Q)$, $I_r=0.5 (I-U)$; see \citealt{hillier_94} for details) -- the definitions of these various quantities are given in Section~2 of \citet{DH11_pol}. The geometry of the axisymmetric ejecta is described with a right-handed set of unit vectors ($\xi_X$ , $\xi_Y$ , $\xi_W$). The origin is the center of the ejecta, the axis of symmetry lies along $\xi_W$, $\xi_Y$ is in the plane of the sky, and the observer lies in the $XW$ plane \citep{hillier_94, hillier_96}. We adopt the same sign convention as in \citet{DH11_pol}. The flux $F_Q$ is positive (negative) when the polarization is parallel (perpendicular) to the symmetry axis. Because of the imposed axial symmetry and the adopted orientation, $F_U$ is zero. Hence, the polarization fluxes that we present from our simulations with \longpol\ will be limited to $F_Q$ and to $P = 100 \times |F_Q| / F_I$ (and thus corresponding to a percentage), where $F_I$ is the total flux. To keep with the convention in spectropolarimetric observations, the normalized  Stokes parameters $q$ and $u$ are defined as $q = F_Q / F_I$ and $u = F_U / F_I$. In this study, we only discuss the flux-like polarized quantities obtained with \longpol. We refer the reader to \citet{DH11_pol} for a discussion of the maps of the polarization intensities on the plane of the sky for various asymmetric configurations and wavelengths. In general, because of the very low level of polarization from asymmetric SN ejecta, these polarization intensity maps are difficult to analyze.

For a \longpol\ calculation, we first construct the 2D ejecta (see Section~\ref{sect_prep_2d_ejecta}). The initial information in this 2D model consists of radius, density, velocity, electron scattering opacity, absorptive opacity as well as emissivity for the selected range of wavelengths. The 2D ejecta is mapped with about one hundred radial points\footnote{This grid is by default the same as in the 1D \cmfgen\ model and thus equally spaced in optical depth scale. However, the two models used may present a recombination front at a different radius. Because the same radial grid is used along all latitudes, it must be augmented to resolve any recombination front along any latitude.} and nine polar angles between pole and equator (equally-spaced; mirror symmetry with respect to the equatorial plane is by default assumed). The radiation field $I$ at each ejecta location ($r,\beta$) is discretized in azimuth $\phi$ and latitude $\theta$. We use thirteen $\phi$ angles while the latitudinal rays are set according to the impact parameter. Fourteen points (or rays) are used to cover uniformly from the ejecta center to the innermost ejecta radius, and the radial grid is used beyond that to define the impact parameter.

In \longpol, most of the computing time is taken by a nested $\beta$ and $\phi$ angle loop (in which the 2D radiative transfer equation is solved for). This section is parallelized with MPI and \longpol\ is run with $N_\beta N_\phi$ processors (hence typically of order a few hundred processors). Convergence is reached after $20-30$ iterations, although the total and polarized fluxes change little after about ten iterations. This convergence is faster and better for optically thin ejecta since the amount of scattering is in that case much reduced. A fully converged model takes approximately 24h with 200 processors (or 400 processors if mirror symmetry is not adopted).

\begin{figure*}[ht!]
\begin{center}
\epsfig{file=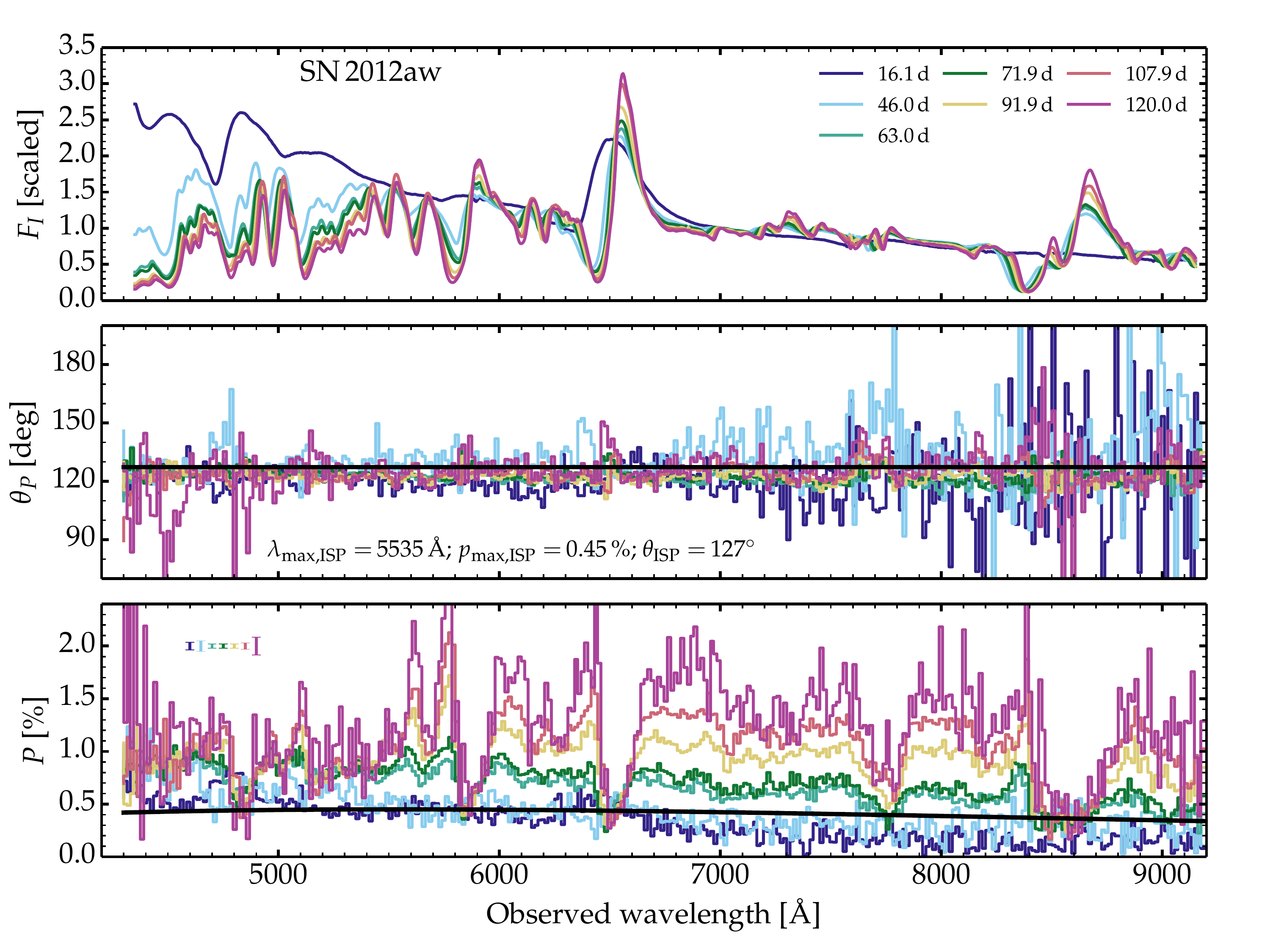, width=17cm}
\end{center}
\vspace{-0.5cm}
\caption{Multiepoch VLT spectropolarimetric observations of SN\,2012aw showing the total flux $F_I$ at the top, the observed polarization angle $\theta_P$ (we show the quantity 180 + $\frac{1}{2}\arctan (F_U,F_Q)$, in degrees), and the observed percentage polarization $P$ (we show here the traditional polarization defined as $100 \sqrt{q^2 + u^2}$; see, e.g., \citealt{leonard_99em_specpol_01}). The quantities $\theta_P$ and $P$ are displayed with a binning of 15\,\AA/bin. Each epoch is color coded (see labels and Table~\ref{tab_obs}). To help gauge the reality of spectropolarimetric features, in the bottom plot we show a representative uncertainty per $15\AA$ bin for each epoch (color coded error bars in the upper left corner), taken as the mean $1\sigma$ uncertainty in the total polarization calculated from 6900\,\AA\ to 7400\,\AA, a region largely free of line features at all epochs. (the particular uncertainty in any given bin rises and falls with the S/N, becoming larger in the P-Cygni troughs and smaller in the peaks.) We overplot the adopted interstellar polarization based on a Serkowski law \citep{serkowski_law_75,cikota_vlt_pol_17} derived by assuming that the intrinsic polarization is zero at the regions of the strong, consistent, depolarization seen near the flux emission peaks of H$\beta$, Na\one\,D, H$\alpha$, and the Ca\two\ near-infrared triplet at epoch 6 (thick black line; see Section~\ref{sect_isp} for details).
\label{fig_obs_only}
}
\end{figure*}

\begin{figure*}[ht!]
\begin{center}
\epsfig{file=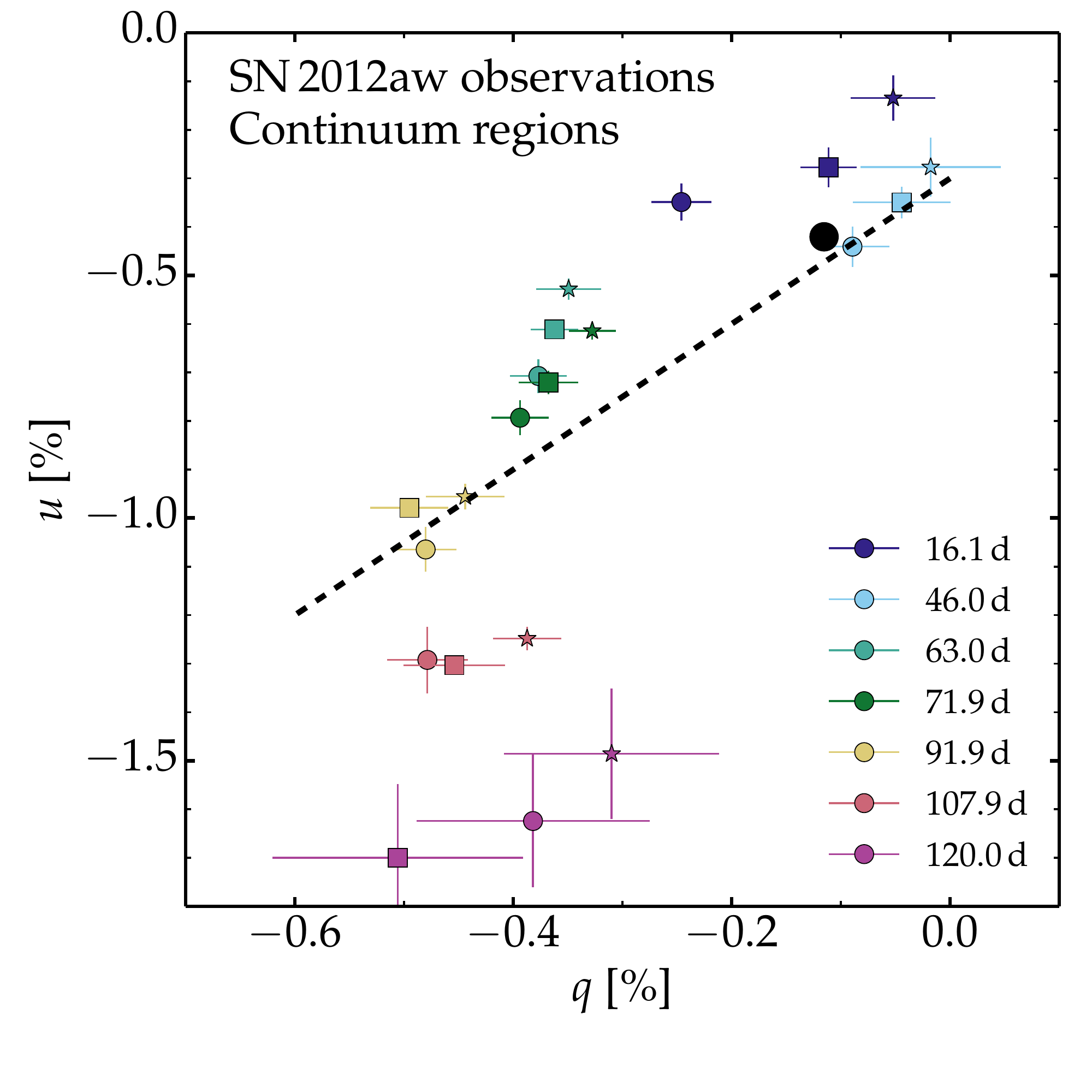, width=8.9cm}
\epsfig{file=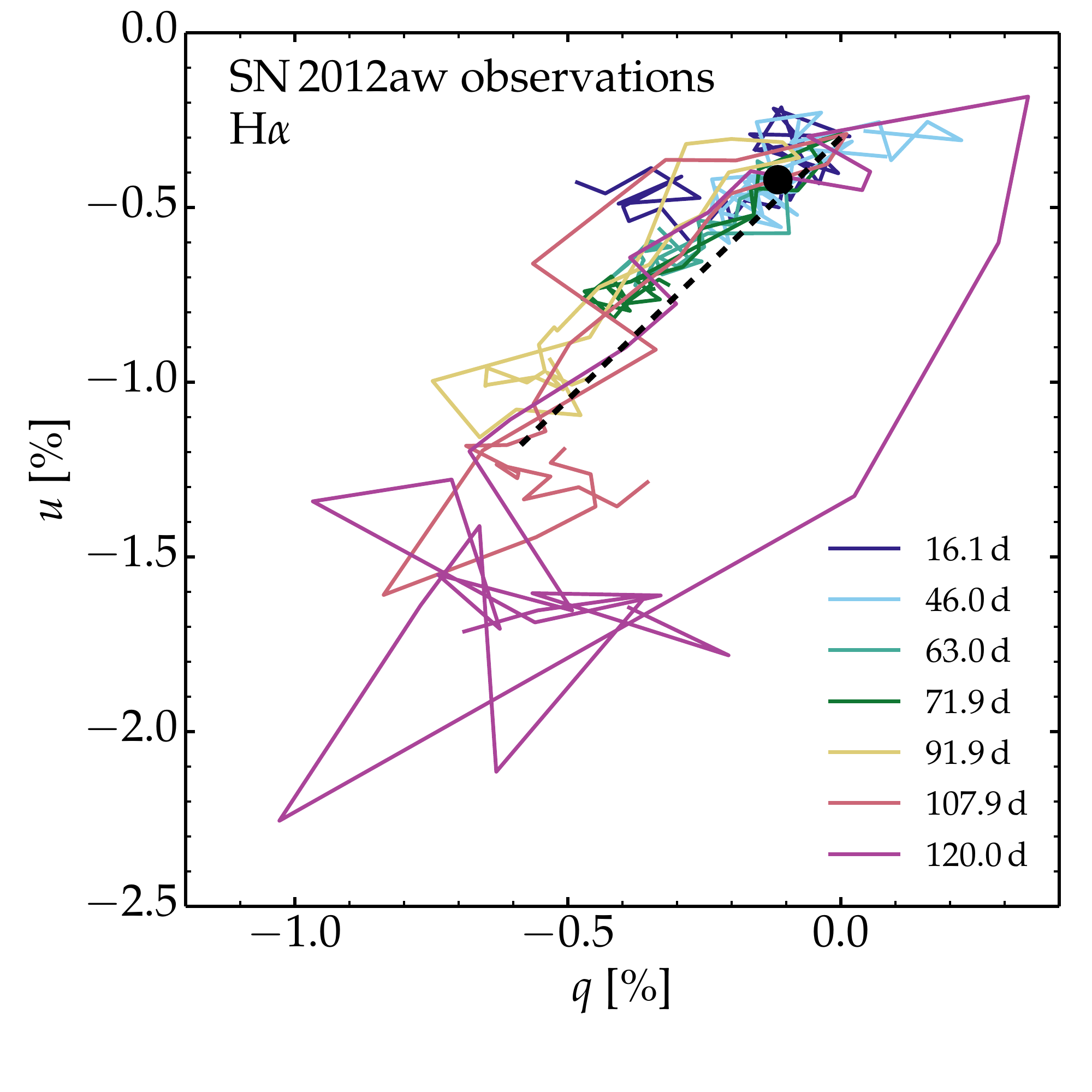, width=8.9cm}
\end{center}
\vspace{-0.5cm}
\caption{Normalized Stokes parameters for SN\,2012aw in three spectral regions devoid of strong lines covering 6000\,$-$\,6150\,\AA\ (circles), 6700\,4$-$\,6900\,\AA\ (square), and 7900\,$-$\,8100\,\AA\ (stars) and for all seven epochs.  $q$ and $u$ were derived by taking the mean of the binned (by 15\,\AA), de-redshifted data over the stated ranges; uncertainties shown are the 1-sigma spread of the data values over the same ranges. Right: Same as left, but now zooming-in on the H$\alpha$ region extending $\pm$\,10,000\,\kms\ away from line center. In both panels, a black dot gives the average interstellar polarization inferred at 6500\,\AA\ and the dashed line indicates the dominant axis adopted for the rotation of the Stokes parameters in subsequent sections.
\label{fig_obs_q_u}
}
\end{figure*}

\section{Observations of SN\,2012\lowercase{aw}}
\label{sect_obs}

\subsection{The VLT\,$-$\,FORS dataset}

We obtained seven epochs of spectropolarimetry of the Type II-P supernova SN\,2012aw using the European Southern Observatory's Very Large Telescope (VLT) at Cerro Paranal in Chile \citep{Appenzeller98}.\footnote{VLT observing program 089.D-0515(A): PI G. Pignata.}  All observations were made with the FOcal Reducer and low dispersion Spectrograph (FORS2), mounted at the Cassegrain focus of the Antu (UT1) telescope.  The data sample days 16 -- 120 following the date of explosion (16.09 March, 2012 UT) estimated by \citet{bose_12aw_13}. Details of the observational epochs, instrumental setup, exposure times, and observational conditions are given in Table~\ref{tab_obs}. The data are presented in Fig.~\ref{fig_obs_only}.

Each complete observational sequence consisted of four separate exposures with the half-wave retarder plate positioned at each of 0, 45, 22.5, and 67.5\,deg; multiple complete sequences were executed at all epochs except for epoch 2, for which only one complete sequence was obtained (see Table~\ref{tab_obs}).  All data were bias subtracted and flat-field corrected using dome flats obtained through the polarimetry optics at a waveplate position angle of 0\,deg on the same nights as the observations\footnote{For epoch 1, the flat fields were acquired on the night after the science observations.}; reductions carried out with unflatfielded data were found to be virtually identical to those made with the flatfielded data, as expected due to the redundant number of half-wave positions obtained \citep[see, e.g.,][]{Patat06}.

One-dimensional sky-subtracted spectra were extracted using both the optimal \citep{Horne86} and standard extraction techniques, with an extraction width set to encompass at least three times the effective ``seeing'' of the particular  observations (see Table~\ref{tab_obs}).  The wide extraction was used to minimize the effect of spurious polarization features being introduced by interpolating the counts in fractional pixels at the edges of narrow extraction apertures (see \citealt{leonard_99em_specpol_01}).  We subtracted a linear interpolation of the median values of background windows on either side of the object from the object's spectrum; for consistency, the ``sky'' background region was set for all extractions to be $\pm\ 7\farcs5:2\farcs5$ from the centroid of the object's spatial profile; an exception was made for the final, seventh epoch, where poor seeing necessitated that a more remote sky region ($\pm\ 8\farcs75:3\farcs75$) be used to assure minimal object light in the sky window.  Each spectrum was then wavelength and flux-calibrated, corrected for continuum atmospheric extinction, and rebinned to a 5~\AA\ bin$^{-1}$ linear scale.

We formed the normalized Stokes $q$ and $u$ parameters and their statistical uncertainties for each complete sequence of observations according to the methods outlined by \citet{1988igbo.conf..157M} and \citet{Cohen97}.  For each observational sequence, we derived results using both the optimal and standard extractions separately, and then compared them to each other. In general, the two extraction algorithms yielded virtually identical results, and so the optimal extractions were preferred due to their (slightly) higher signal-to-noise ratio and ability to discount minor cosmic ray strikes. However, in certain spectral regions that contained particularly sharp spectral features that also affected the ``sky'' region (e.g., regions of strong telluric absorption or, more rarely, severe cosmic-ray strikes across the spatial profile of the object on the CCD), the optimal extraction algorithm produced spurious features in the polarimetry that were not found in the standard extraction results (and, were not consistent from one observational set to the next on the same night).  Presumably, such features occur due to a breakdown of the assumption of a ``smoothly varying'' spatial profile for the object, which underlies the successful application of the optimal weighting algorithm \citep{Horne86}. This occurred quite frequently at the telluric ``A-band'' feature near $7600$ \AA; in one instance, it also occurred in the vicinity of a particularly bad cosmic-ray strike.  In these limited regions (typically $\lesssim 40$ \AA), the standard extraction's polarimetric results were inserted; otherwise, the results derived from the optimal algorithm were used.

The fact that multiple, identically obtained, complete observational sequences were taken at nearly all epochs (see Table~\ref{tab_obs}) permitted us to perform valuable tests on the reduced data and calculated uncertainties. To build confidence in our calculated statistical uncertainties, we utilized the data for epoch 3, for which 13 complete, independent, and identically obtained sets were acquired, and subjected them to the following test. First, we computed for each $5 \AA$ wavelength bin the $1\sigma$ spread of values from all 13 sequences for a Stokes parameter (we took the normalized Stokes $q$); in principle, this should give a rough estimate of the $1\sigma$ uncertainty for each single set of data. (The calculated mean uncertainty derived in this way across the entire spectrum was $\sim 0.25\%$ per bin.) We then compared this with the statistical uncertainty computed through our algorithm for a single observational sequence, and found the two estimates to agree with high precision: The mean difference between the two calculations was less than $0.01\%$ across the spectrum. With confidence in our calculated statistical uncertainties, then, we performed a final check on the veracity of all statistically significant features seen in the polarimetry at all epochs (except epoch 2, for which only one complete set was obtained; see Table~\ref{tab_obs}) by intercomparing the individual sequences' results, carefully examining each one for spurious features not seen in the others.

We then corrected all data for the small ($\lesssim 0.1\%$) amount of instrumental polarization known to exist at the VLT \citep{cikota_vlt_pol_17}. To correct for the wavelength dependence of the measured polarization angle induced by chromatism of the retarder plates, we applied the chromatic zero angle correction provided by the VLT website\footnote{\url{http://www.eso.org/instruments/fors/inst/pola.html}} to all data. In addition, we checked the absolute zero-point of the derived polarization angle in the sky coordinate system on each night by observing polarized standard stars from the lists of  \citeauthor{Clemens90} (1990; HD~127769) and \citeauthor{cikota_vlt_pol_17} (2017; Hiltner652; BD-144922; HDE316232; and Vela1). With the exception of the final (seventh) epoch, all polarized standard stars' derived polarization angles (in the $V$-band; nominal wavelength range of 5050\,$-$\,5950 \AA) agreed with the cataloged  values to within 0.6\,deg.  On the seventh epoch, the polarization angle derived for HD~127769 differed from the value reported by \citet{Clemens90} by 2.4\,deg; the cause for this discrepancy is not known, although we do note the unusually poor seeing (Table~\ref{tab_obs}) and short exposure time for the star (one second per waveplate position; more typical of other nights was 3\,$-$\,5 seconds) on this night. In all cases, including this final epoch, we kept the nominal polarization angle derived directly from the data, and did not apply any overall zero-point offset based on our observations of polarized standard stars. To check for instrumental polarization beyond that already accounted for through application of the linear relations provided by \citet{cikota_vlt_pol_17}, we also observed the null polarization standard HD~109055 \citep{Clemens90,Berdyugin95} at every epoch, and always found it to be null to within $0.05\%$.

For each epoch, the normalized Stokes $q$ and $u$ parameters derived from all complete observational sequences were combined according to their statistical weights, with new uncertainties calculated. We then removed the interstellar polarization derived in Section~4.2, removed a redshift of 778\,\kms\ \citep[][via NED]{Devaucouleurs91}, rebinned to $5$~\AA\ bin$^{-1}$, and created a final, rest-frame dataset with a wavelength range of $4350$ \AA\ -- $9150$ \AA.  We also computed results binned to 15, 25, 50 and 100 \AA\ bin$^{-1}$ for presentation purposes.  To generate the total flux spectrum for each epoch, a final correction for telluric absorption bands \citep{Wade88} was made.

To illustrate the continuum polarization as a function of time we show in the left panel of Fig.~\ref{fig_obs_q_u} the normalized Stokes $q$ and $u$ measurements for all seven epochs in three spectral regions devoid of strong lines around 6000, 7000, and 8000\,\AA. All measurements lie in a single quadrant of the $(q,u)$ plane, and to first order lie along a straight line. The biggest departure from a straight line occurs for the data at 120\,d. The offset is significant, although the data at this epoch also has the worst signal-to-noise ratio. If SN 2012aw were axisymmetric all polarization data would lie along a straight line in the $(q,u)$ plane, and, in the absence of interstellar polarization, would pass through the origin. The presence of a dominant axis in the data sets indicates that SN 2012aw must possess a strong degree of axial symmetry. However, the scatter (and in some cases one could argue for an epoch dependent offset) of the polarization observations about the straight line indicate the structure cannot be simply axisymmetric.

A study of the polarization behavior for H$\alpha$ (right panel of Figure~\ref{fig_obs_q_u}) yields a similar conclusion to that reached using the continuum polarization.  Again, all measurements lie in a single quadrant of the $(q,u)$ plane, and to first order lie along a straight line. As for the continuum, it is the observations at 120\,d that show the largest offset. Because of the presence of a dominant axis, we rotate the observations so that the bulk of the polarization lies in $q$ (Section~\ref{sect_rotate}). This has the consequence that sign reversals in $q$ can easily be seen in the same plot, whereas such reversals are not seen in $P$ and imply a 90-deg change in the angle of polarization.

\subsection{Interstellar polarization}
\label{sect_isp}

Correcting for interstellar polarization is a delicate problem with no perfect solution. In this study, we attempted two different techniques that rely on quite different theoretical assumptions.

Our first approach was to assume that the intrinsic polarization of the SN is zero at the first epoch; thus, all observed polarization should be due to the ISP. Theoretically, this may be motivated by the fact that instabilities, born in the progenitor core during the explosion or at the H/He interface after shock passage, influence little the outer ejecta. Unfortunately, there is clearly some variation in polarization across lines even at the earliest epoch, incompatible with the expected smooth featureless polarization from the interstellar medium. Nonetheless, we fit a spline to the ``continuum'' regions (i.e., those away from strong line features) as a first estimate of the ISP.

For our second approach we considered the theoretical prediction that during the optically thick, ``photospheric'' phase of an SN\,II's development (up through about day $\sim$\,115 for SN\,2012aw, when its light curve begins to drop off of the plateau; \citealt{bose_12aw_13}), broad, unblended emission features (e.g., H$\alpha$, the Ca\two\ near-infrared triplet) may possess zero intrinsic SN polarization \citep[e.g.,][]{trammell_specpol_93,hoeflich_pol_96,DH11_pol}. If complete depolarization of all SN light (line and continuum) has occurred in such spectral regions --- a further assumption commonly employed to estimate the total ISP in SN studies \citep[e.g.,][]{howell_pol_01,leonard_04dj_06,nagao_17gmr_pol_19} as well as polarization studies of emission-line stars \citep[e.g.,][]{meyer_pol_02} --- then any measured  polarization there is due solely to ISP. For multiepoch data, this should present as fixed, unchanging levels of polarization (and, polarization angle) with time in these spectral regions, since ISP does not change appreciably over the timescales involved whereas intrinsic SN polarization may.

Examination of Fig.~\ref{fig_obs_only} shows that the expectation is quite well realized empirically for SN\,2012aw, where sharp, consistent, depolarizations are seen for several strong lines, including H$\alpha$, H$\beta$, Na\one\,D, and the Ca\two\, near-infrared triplet. That is, while nearby regions of the spectrum vary by up to $\Delta P \sim$\,1.5\,\% through the observational sequence, these spectral regions remain consistent to within $\Delta P \sim$\,0.1\,\%. Motivated by this consistency, we fit a Serkowski law \citep{serkowski_law_75} through the strong depolarizations seen in H$\alpha$, H$\beta$, Na\one\,D, and the Ca\two\ near-infrared triplet to the epoch 6 data (similar results were found using epochs 3 to 7). Using these four spectral locations as a reference for the interstellar polarization, we searched for the set of values for $\lambda_{\rm\max,ISP}$, $q_{\rm max,ISP}$, and $u_{\rm max,ISP}$ that produced the best fitting curve given by a Serkowski law. Each $q_{\rm max,ISP}, {\rm\ and\ } u_{\rm max,ISP}$ correspond to $\theta_{\rm ISP}$ and $p_{\rm max,ISP}$ through

\begin{equation}
 \theta_{\rm ISP} = 0.5\arctan(u_{\rm max, ISP} / q_{\rm max, ISP})
\end{equation}
and
\begin{equation}
p_{\rm max, ISP} = q_{\rm max, ISP}\cos(2 \theta_{\rm ISP}) +
                                       u_{\rm max, ISP}\sin(2 \theta_{\rm ISP}).
\end{equation}

The wavelength dependence of the interstellar polarization is then given by
\begin{equation}
p_{\rm ISP}(\lambda) = p_{\rm max, ISP} \, \exp \big(-K_{\rm ISP}\ln^2 (\lambda_{\rm max, ISP}/\lambda) \big),
\end{equation}
where $K_{\rm ISP} = 1.13 + 0.000405 \lambda_{\rm max,ISP}$ \citep{cikota_pol_18}, with $\lambda_{\rm max,ISP}$  given in Angstrom units. With this approach, the best fitting values for SN 2012aw are $\lambda_{\rm max,ISP} = 5535 \AA, q_{\rm max,ISP} = -0.12\%, u_{\rm max,ISP} = -0.43\%$, or $p_{\rm max,ISP} = 0.45\%, \theta_{\rm ISP} = 127^\circ$.  We then obtain
\begin{equation}
q_{\rm ISP}(\lambda) = p_{\rm ISP}(\lambda) \cos(2 \theta_{\rm ISP}) \,  ; \,
u_{\rm ISP}(\lambda) = p_{\rm ISP}(\lambda) \sin(2 \theta_{\rm ISP}).
\end{equation}

While both estimates of the ISP have weaknesses, we encouragingly found them to agree to better than $\sim 0.2\%$ over the full spectral range covered by our observations. We ultimately adopted the ISP described by our second approach and proceeded to correct all spectropolarimetric observations of SN 2012aw by it. However, given the difference between our estimates, we expressly recognize that a $\sim$\,0.2\,\% systematic uncertainty in the absolute level of all calculated ``intrinsic'' polarizations exists in all that follows.\footnote{Given this uncertainty in the ISP, one should not over-interpret, for example, the presence of small sign flips in the data.} Fortunately, since the ISP is a smooth function of wavelength, this uncertainty in its absolute value has no effect on interpretation of specific line features, which is the main focus of this work. Further, as the degree of continuum polarization ultimately seen in SN\,2012aw approaches $\sim$\,1.2\,\% in the latest epochs, the effects of a 0.2\,\% uncertainty do not qualitatively alter our fundamental conclusions regarding the overall polarization level when comparing to our models.

Adopting this interstellar polarization and focusing on the spectral range between 6900 and 7200\,\AA\  (Fig.~\ref{fig_obs_only}), SN\,2012aw yields a nonzero polarization even at the earliest times, although interpretation of this is tempered by the fact that its value, $\sim$\,0.2\,\%, is of order our systematic uncertainty in the ISP. SN\,2012aw then shows a strong temporal increase in polarization during the second half of the plateau phase, reaching a maximum of about $1.2\%$ as the SN starts plunging in brightness and transitioning to the nebular phase. This has been seen in SN 2004dj \citep{leonard_04dj_06}, SNe 2006ov and 2007aa \citep{chornock_pol_10}, SN 2008bk \citep{leonard_08bk_12}, and 2013ej \citep{leonard_iauga_15}. SN 2012aw could not be observed at nebular times as it went behind the sun. It was too faint when it later emerged.

\begin{figure*}
\begin{center}
\epsfig{file=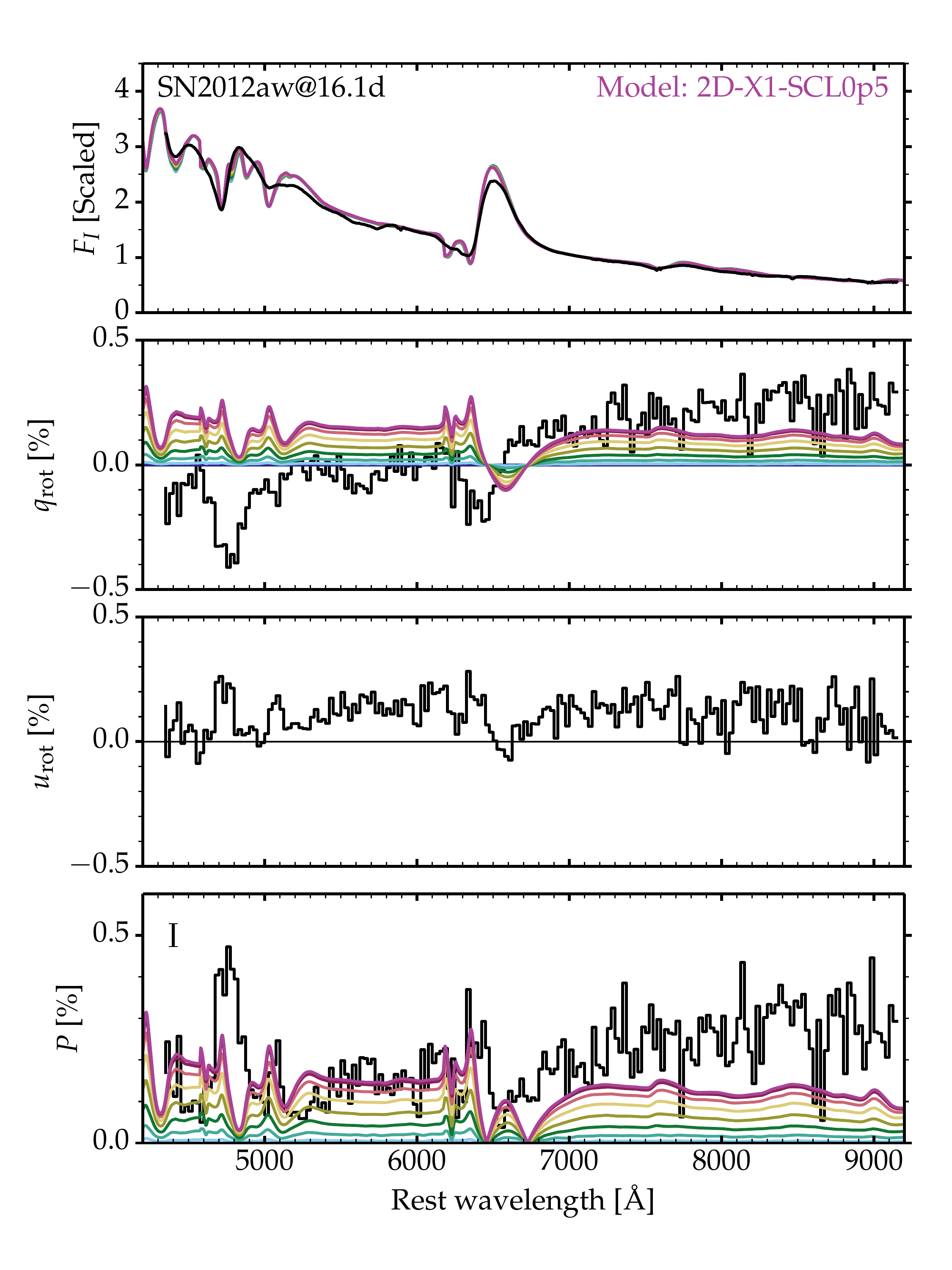, width=8.5cm}
\epsfig{file=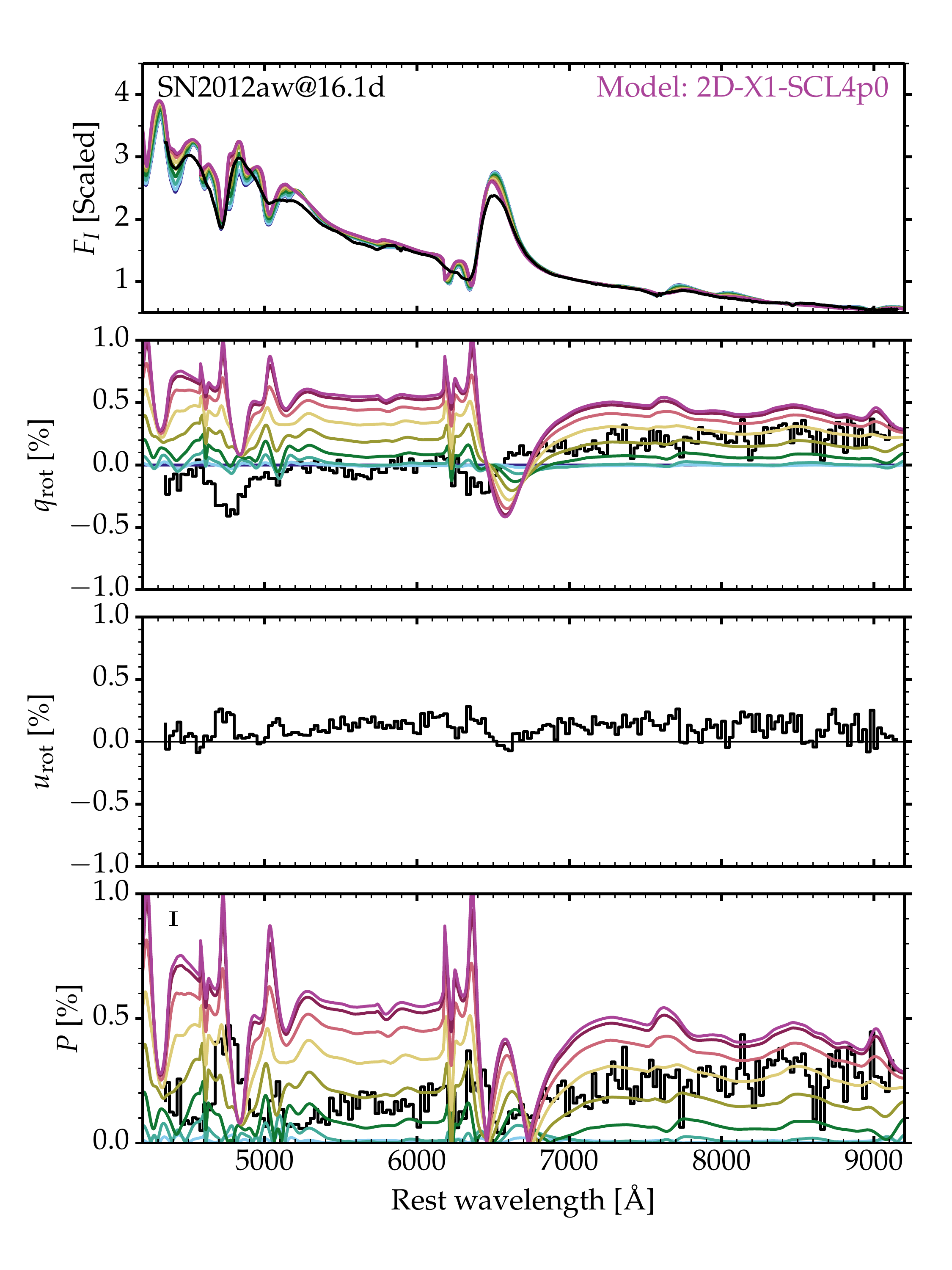, width=8.5cm}
\end{center}
\vspace{-0.5cm}
\caption{Left: Comparison between the model 2D-X1-SCL0p5 and observations of SN\,2012aw for epoch 1 (i.e., 16.1\,d after explosion) using a density scaling with $A$ of 0.5 (left) or 4.0 (right). All inclinations between polar and equatorial directions are shown. For SN\,2012aw, we correct for the instrumental polarization, for the interstellar polarization and redshift (as well as reddening for $F_I$). The observed normalized Stokes parameters have been rotated by 28\,deg (see section~\ref{sect_rotate}). The ordinate range is the same for $q_{\rm rot}$ and $u_{\rm rot}$. A binning of 25\,\AA\ is used for $q_{\rm rot}$, $u_{\rm rot}$, and $P$, while a binning of 5\,\AA\ is used for $F_I$. As described in the caption to Fig.~\ref{fig_obs_only}, a representative $1\sigma$ uncertainty in the polarization was calculated, and is shown in the bottom panels. The polarization displayed in the bottom panels for the SN\,2012aw data is the ``traditional polarization'', as defined in the caption to Fig.~\ref{fig_obs_only}. For all models $u_{\rm rot}$ is null across the spectrum, and so is not plotted.
\label{fig_pol_scale}
}
\end{figure*}

\subsection{Rotation of the Stokes fluxes}
\label{sect_rotate}

For an  axially symmetric ejecta the polarization (at all times and at all wavelengths) lies on a diagonal line in the $(q,u)$ plane which passes through the origin. Since the slope of this line simply reflects the orientation of the ejecta symmetry axis on the plane of the sky, one can rotate it so that all the polarization lies in $q$. This then allows a direct comparison with our model calculations in which all the polarization lies in $q$. In observations, noise and departures from strict axial symmetry leave a residual signal in $u$ even after such a rotation has been applied.

Hence, when we compare to our polarization results for axisymmetric models, we apply a rotation of the observed Stokes vectors by an angle  $\theta_{\rm rot} = 0.5 \arctan (\delta u / \delta q) $, where  $\delta u$ and  $\delta q$ are measured in the $(q,u)$ plane. This is done to put most of the observed polarization in $q$. The rotated fluxes are then given by
\beq
q_{\rm rot} =   q \cos(2 \theta_{\rm rot}) + u \sin (2\theta_{\rm rot})
\eeq
and
\beq
u_{\rm rot} = -q \sin(2 \theta_{\rm rot}) + u \cos (2\theta_{\rm rot}).
\eeq

The data of SN\,2012aw, corrected for interstellar polarization, yield an observed polarization angle of 118\,deg in the plane of the sky. However, we choose to rotate the Stokes vectors by the angle orthogonal to this -- that is, of 28\,deg -- in order to produce a sign of the rotated Stokes vectors that agrees with the models, for ease of comparison (rotating through an angle 90\,deg different simply changes the sign of the polarization).  With this rotation, the bulk of the polarization is now in $q$, and $u$ is generally close to zero. This works with only modest success at early times when the polarization level is low, but it seems suitable for epochs four and later, as shown in the next section.

\begin{figure*}
\begin{center}
\epsfig{file=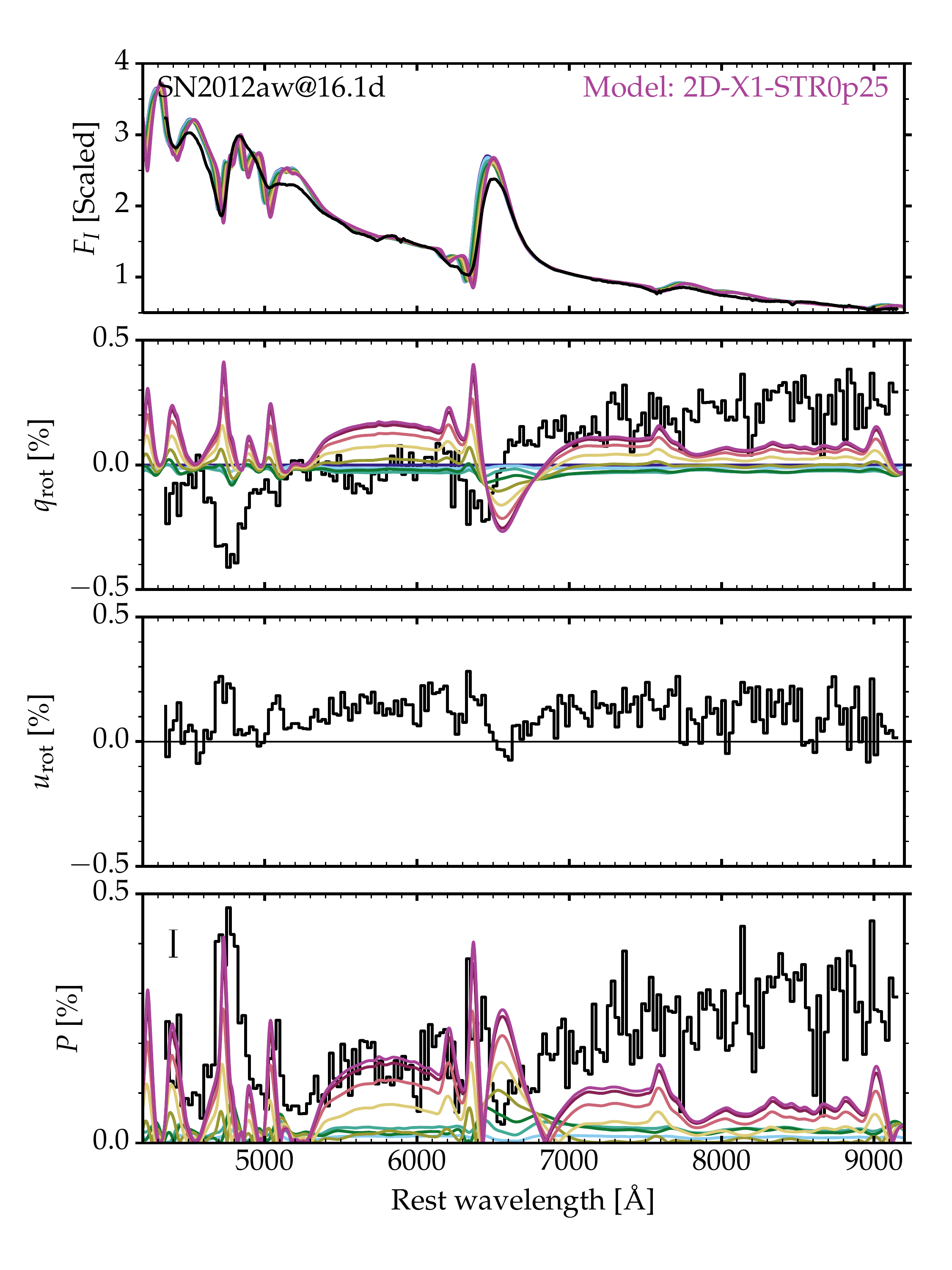, width=8.5cm}
\epsfig{file=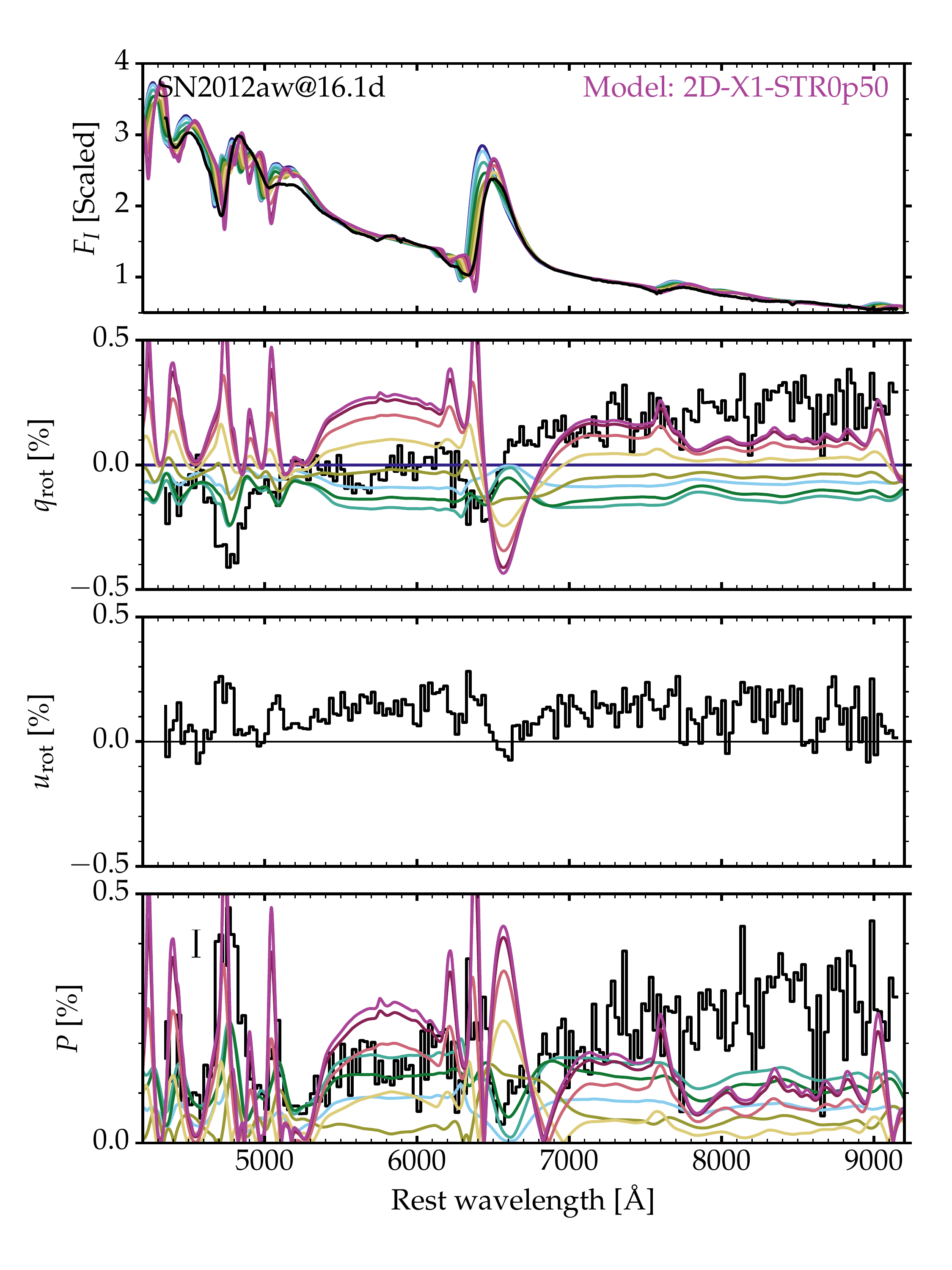, width=8.5cm}
\end{center}
\vspace{-0.5cm}
\caption{Same as Fig.~\ref{fig_pol_scale}, but now for the 2D ejecta models built using a stretching of model 1D-X1, with a parameter $A$ of 0.25 (left; model 2D-X1-STR0p25) and 0.50 (right; model 2D-X1-STR0p50).
\label{fig_pol_stretch}
}
\end{figure*}

\section{Modeling results and comparisons to observations of SN\,2012\lowercase{aw}}
\label{sect_res}

As discussed in Section~\ref{sect_prep_2d_ejecta}, we used three different ways to introduce the asymmetry in our 2D ejecta. In the first two sections below, we describe the results for the first epoch adopting a latitudinal density scaling (Section~\ref{sect_prep_lat_scl}) or a latitudinal stretching (Section~\ref{sect_prep_lat_str}) of a 1D \cmfgen\ model. In the subsequent section, we use  the combination of two models (Section~\ref{sect_prep_lat_multi}) and compare to all observed epochs of SN\,2012aw (Section~\ref{sect_obs}). For clarity, we call these models the scaled model, the stretched model, and the hybrid model.

Throughout this section, we present multipanel figures that have the same structure. The top panel shows the total flux $F_I$ for both the model  and the observations (normalized to unity at 7100\,\AA). The next two panels show the normalized, rotated Stokes parameters $q_{\rm rot}$ and $u_{\rm rot}$ (we plot the percentage value), also corrected for interstellar polarization (see previous section). Finally, the bottom panel shows the  percentage polarization $P$ for both the observation (black) and the model (colors other than black). The model results are sometimes shown for all inclinations. The inclination of zero degree yields zero polarization. The other inclinations are spaced uniformly every 11.25\,deg, colored from blue to red (colorblind rainbow colormap) as the inclination is increased.

Compared with the simulations presented in \citet{DH11_pol}, the present simulations provide the total and polarized flux spectrum over the entire optical range. We explicitly account for line overlap or blanketing and can directly compare to the optical VLT\,$-$\,FORS spectropolarimetric data that we collected for SN\,2012aw.

\subsection{Results with the latitudinal density scaling at early times}
\label{sect_scl}

Figure~\ref{fig_pol_scale} compares the observations of SN\,2012aw at the first epoch (16.1\,d after explosion) with models in which a latitudinal density scaling has been applied to the 1D \cmfgen\ model 1D-X1.  The left panel shows the results for $A=0.5$ and the right panel for $A=4.0$ (see Eq.~\ref{eq_rho_lat_scaling}), corresponding to a density contrast at a given radius of 1.5 and 5.0 between pole and equator (prolate density configurations). The simulations assume mirror symmetry with respect to the equatorial plane. For both models, the asymmetry introduces only a minor change with inclination for the total flux $F_I$ (top panel). The polarized flux $F_Q$ (or equivalently $q$) is positive throughout most of the optical range, which means that the electric vector is parallel to the symmetry axis. This is an optical depth effect because in the optically-thin regime, $F_Q$ would be negative in this case. Switching to an oblate configuration would result in a sign change of the polarization (rotation of the polarization angle by 90\,deg). In absolute terms, the polarized flux is here greater where the flux is greater.  The percentage polarization is also greater at shorter wavelengths, albeit modestly. Theoretically, one may expect a greater polarization at shorter wavelength because of the greater albedo toward the Balmer edge \citep{DH11_pol}, especially at this young SN age when the material in the spectrum formation region is at least partially ionized. The polarization changes between lines and continuum, as well as within the lines. We see a sign reversal of the polarization in H$\alpha$, and for many lines in the case of a stronger asymmetry (right panel of Fig.~\ref{fig_pol_scale}). For a large scaling ($A=4.0$), the model $P$ rises to a maximum of 0.6\% in the largely line-free region between Fe\two\,5169\,\AA\ and H$\alpha$.

The observed polarization of SN\,2012aw at the corresponding epoch differs in a number of ways from the model predictions. It is greater at redder wavelengths, although at this epoch the polarization is quite low, and only about 0.2\% at 7000\,\AA, thus significantly lower that the maximum value obtained with the models discussed above (about 0.6\%). Observations also reveal a sizable change in polarization across H$\beta$ and H$\alpha$, but not other lines (perhaps because the data are too noisy). While H$\beta$ and H$\alpha$ in the models also show polarization, the structure does not match the observations.  In $q$ the model shows an enhanced polarization in the P~Cygni absorption component (not seen in the observations), and the polarization absorption trough is offset from that observed. In $P,$ the models show a very complicated structure which contrasts with the relatively smooth structure seen in the observations. The difficulty in interpreting this epoch is exacerbated by the uncertainties in the interstellar polarization,  the low level of polarization, and the significant signal that remains in the rotated $u$ parameter.

\begin{figure}
\begin{center}
\epsfig{file=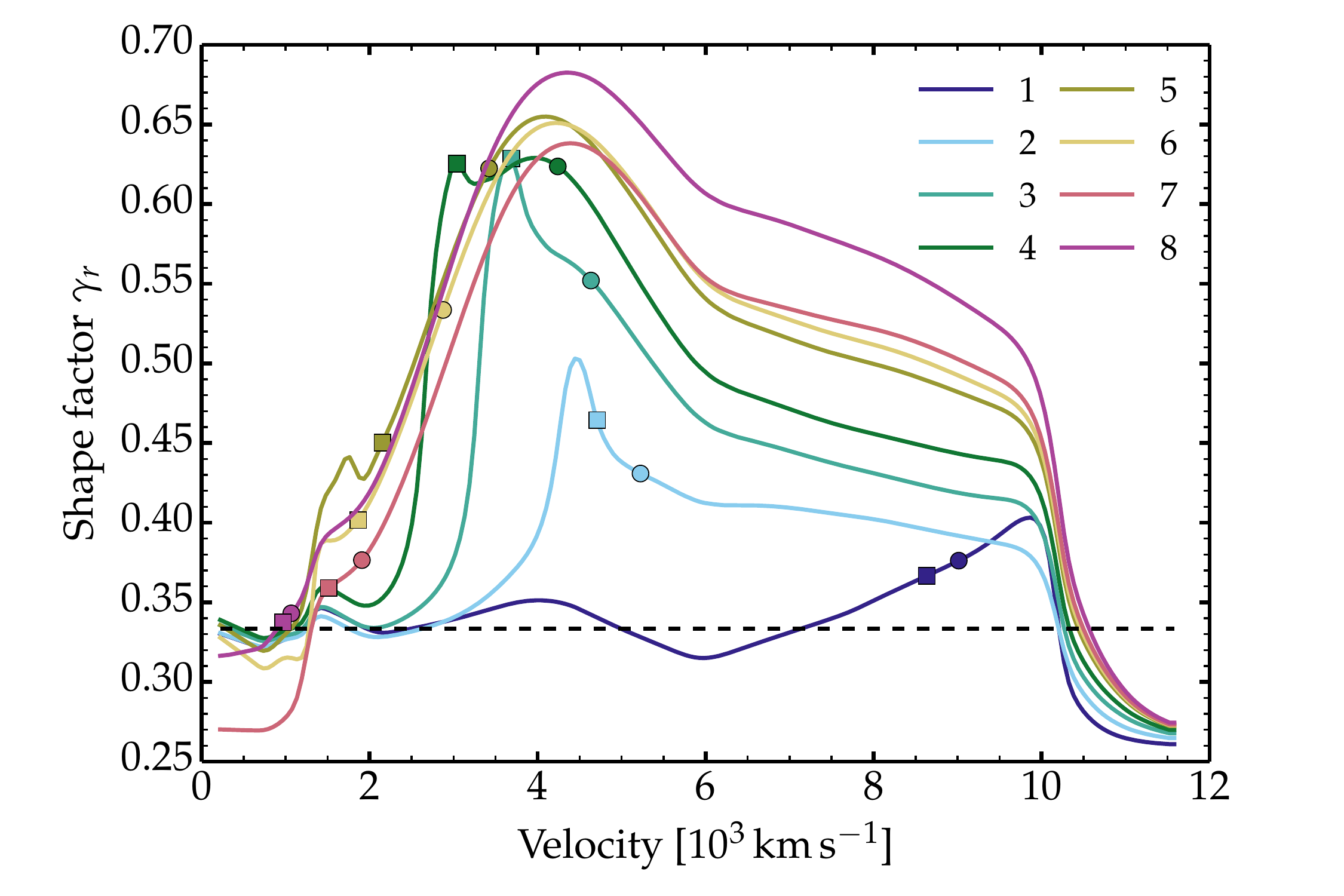, width=9.2cm}
\end{center}
\vspace{-0.5cm}
\caption{Evolution of the shape factor $\gamma(r)$ in the hybrid model versus velocity for the eight epochs modeled (the last epoch at 160\,d has no observational counterpart). The symbols correspond to the photospheric radii in the equatorial direction (square) and the polar direction (dot).
\label{fig_gamma_r}
}
\end{figure}

\subsection{Results with the latitudinal stretching at early times}
\label{sect_str}

For an asymmetric ejecta resulting from a radial stretching of the 1D \cmfgen\ model 1D-X1 (see section~\ref{sect_prep_lat_str} for explanations), the results are quite different from models with a latitudinal density scaling (Fig.~\ref{fig_pol_stretch}). The changes to the total flux are greater than for the models with the density scaling. The flux $F_I$ at 7100\,\AA\ increases by 40\% (90\%) as we vary the inclination from 0 to 90\,deg in model  2D-X1-STR0p25 (2D-X1-STR0p50; see Section~\ref{sect_prep_lat_str} for details on the nomenclature). Once we renormalize the spectra at 7100\,\AA, the variation with inclination appears mostly in lines in the form of a wavelength shift. This shift is more noticeable in model 2D-X1-STR0p50. The polarization is also overall lower. Varying the magnitude of the radial stretching changes little the resulting polarization, despite the large density contrast between pole and equator (corresponding to values of 14.6 and 130.0 for models 2D-X1-STR0p25 and 2D-X1-STR0p50 since the ejecta follow a power-law density with exponent twelve in the outer regions). In both cases, we see strong sign reversals of the continuum polarization when the inclination is changed. There are also strong reversals of the polarization across line profiles for inclinations corresponding to low latitudes (near equator-on views). Such signatures are likely due to optical depth effects with a complicated cancellation of the local polarization on the plane of the sky. These can occur at early times because asymmetry impacts the distribution on the plane of the sky of both the scatterers and the flux. Hence, for the case of a latitudinal stretching, the model fails to reproduce the greater polarization at longer wavelengths, is unable to reproduce the level of polarization observed, and fails to match the structure across H$\alpha$.

\begin{figure*}[h!]
\begin{center}
\epsfig{file=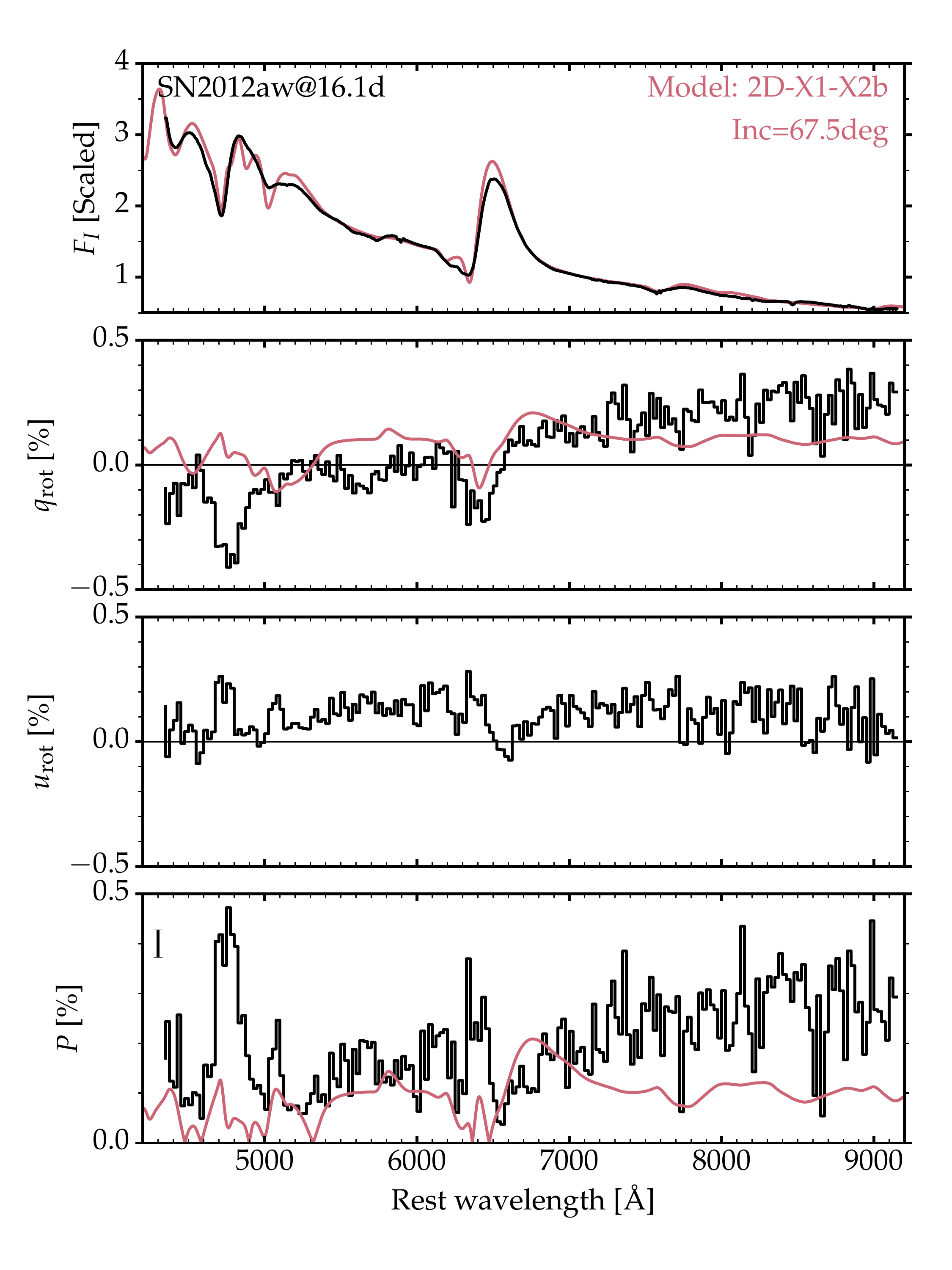, width=8.cm}
\epsfig{file=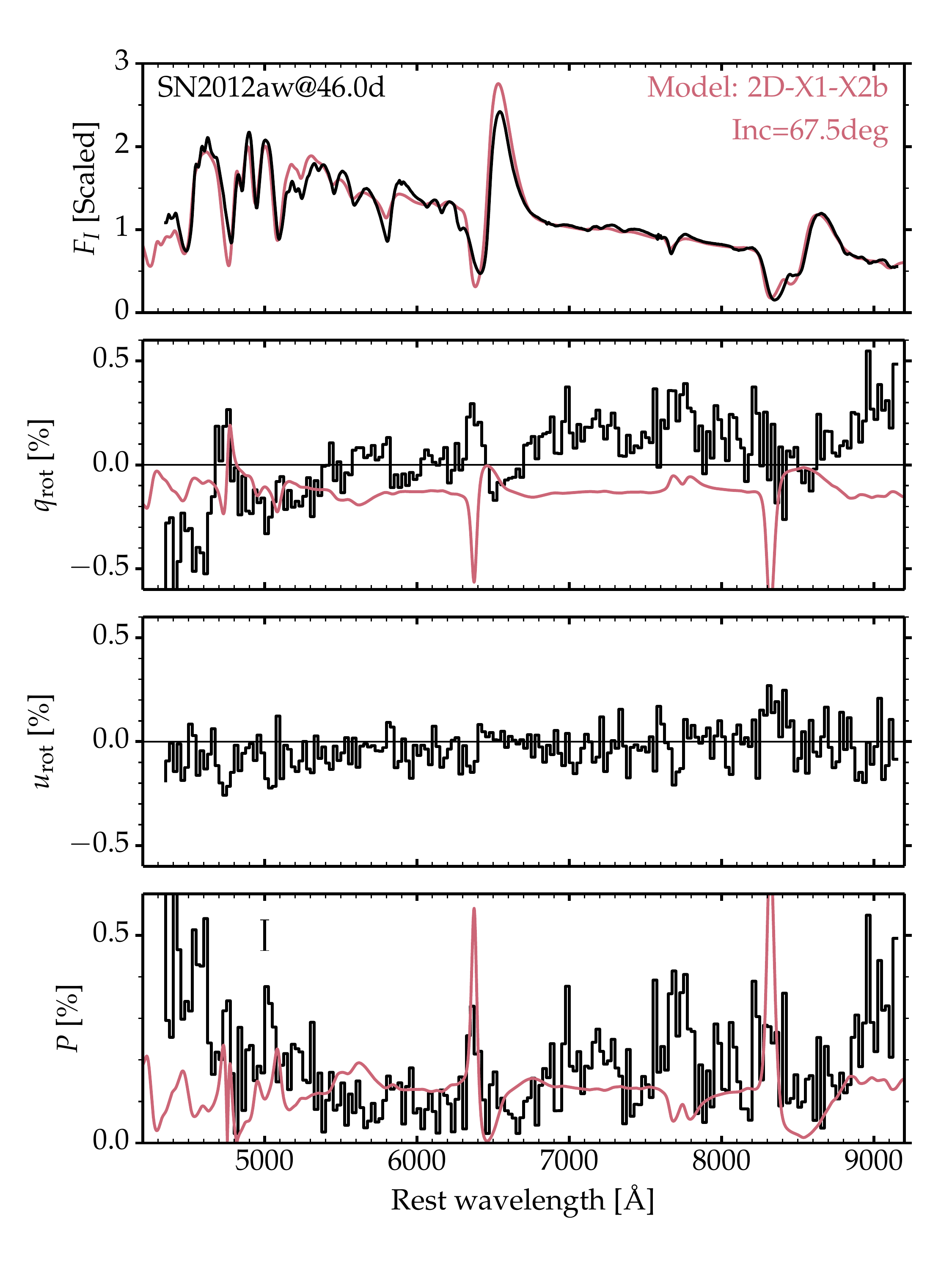, width=8.cm}
\epsfig{file=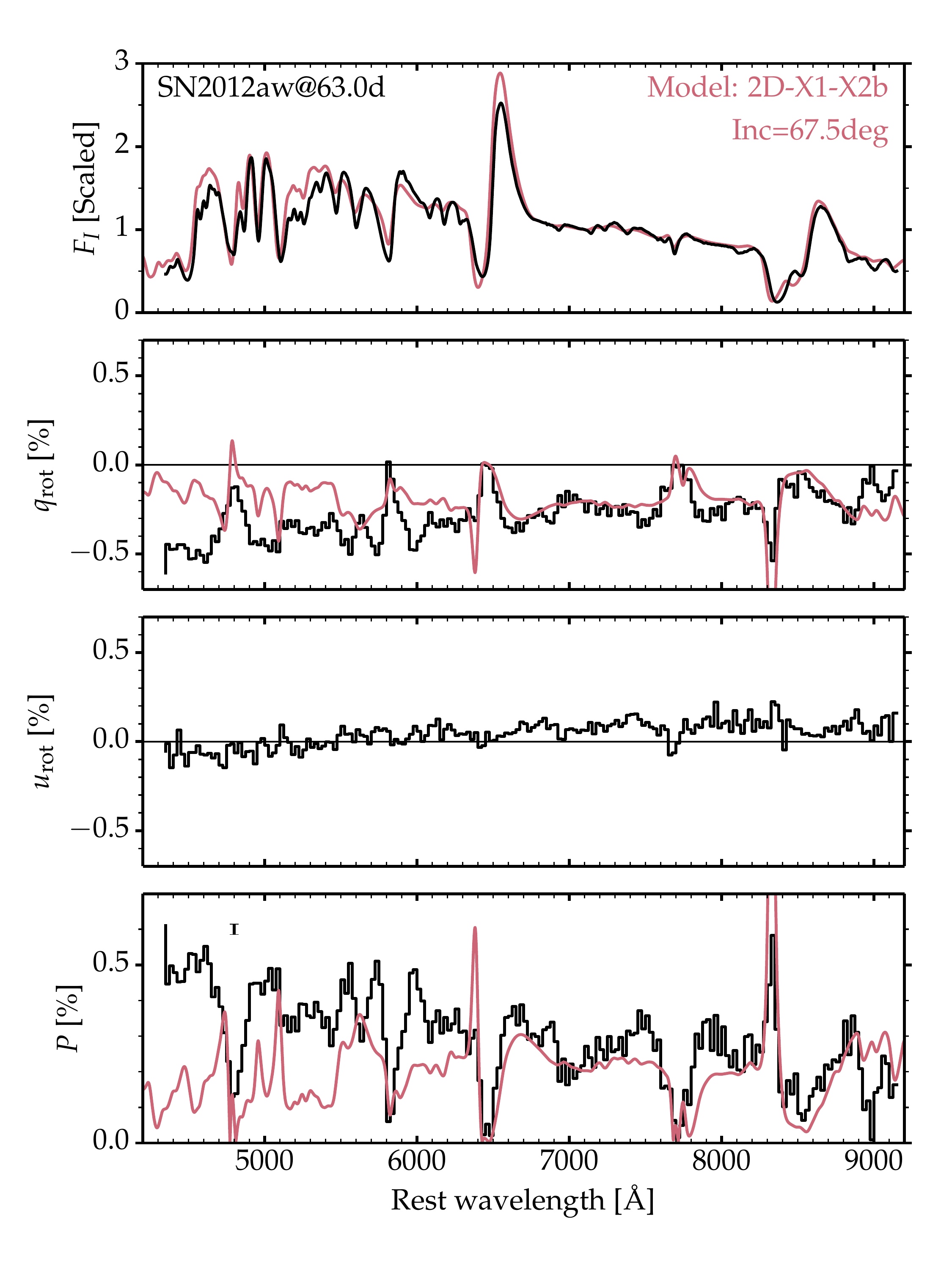, width=8.cm}
\epsfig{file=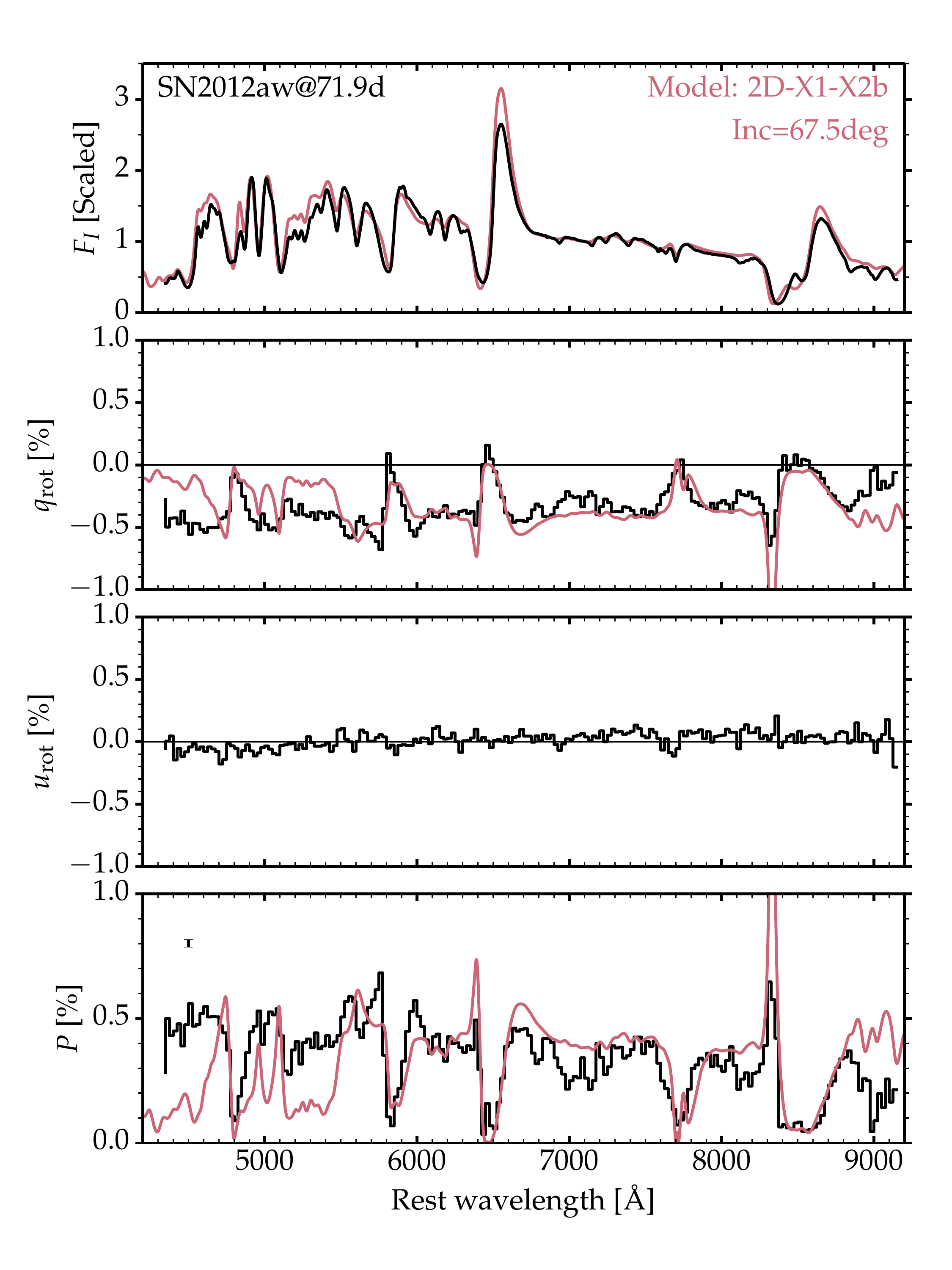, width=8.cm}
\end{center}
\vspace{-0.5cm}
\caption{Same as Fig.~\ref{fig_pol_scale}, but now for the hybrid model 2D-X1-X2b (for an inclination of 67.5\,deg) and the observations of SN\,2012aw for epochs 1, 2, 3, and 4.
\label{fig_pol_1}
}
\end{figure*}

\begin{figure*}[h!]
\begin{center}
\epsfig{file=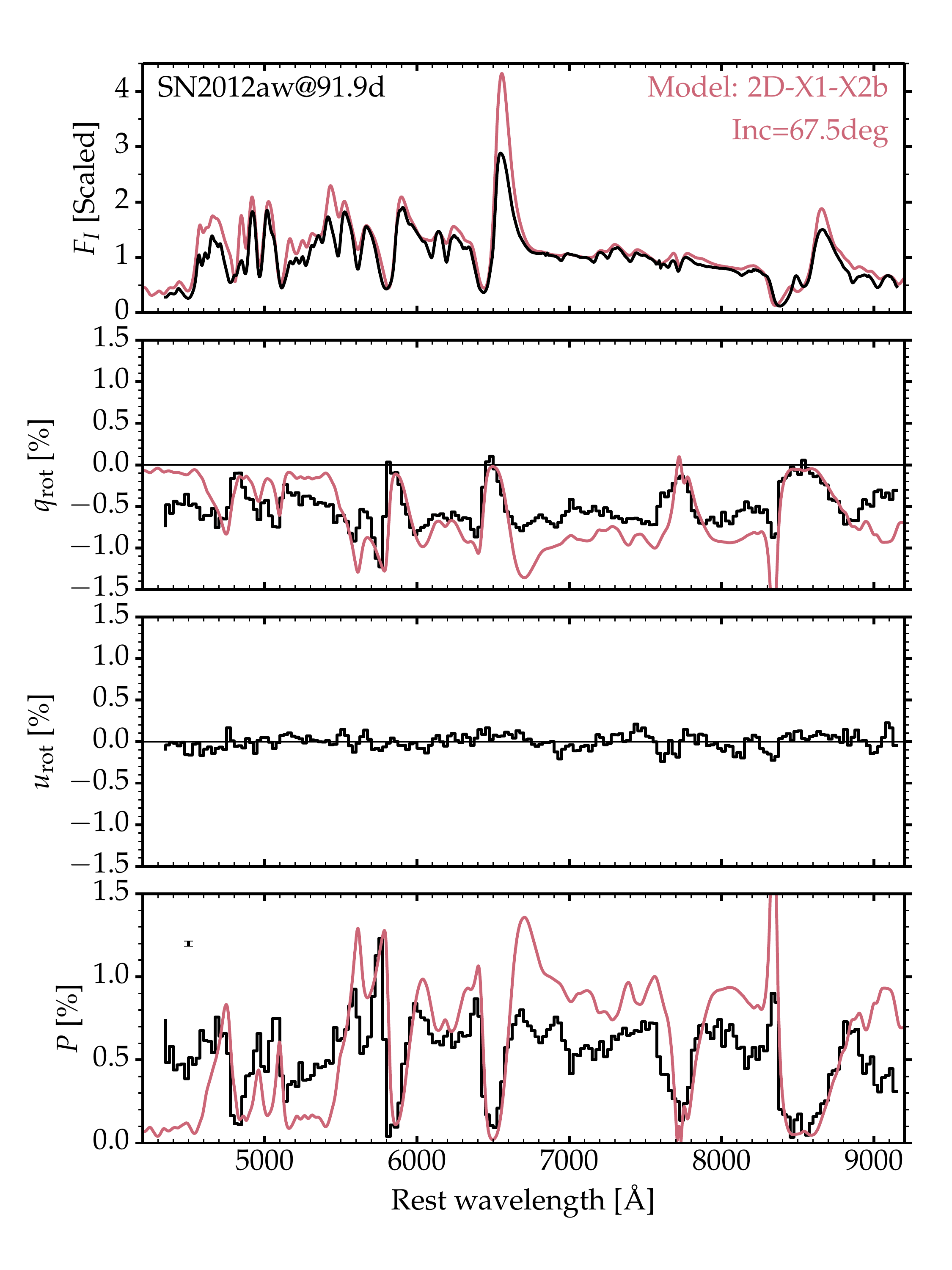, width=8.cm}
\epsfig{file=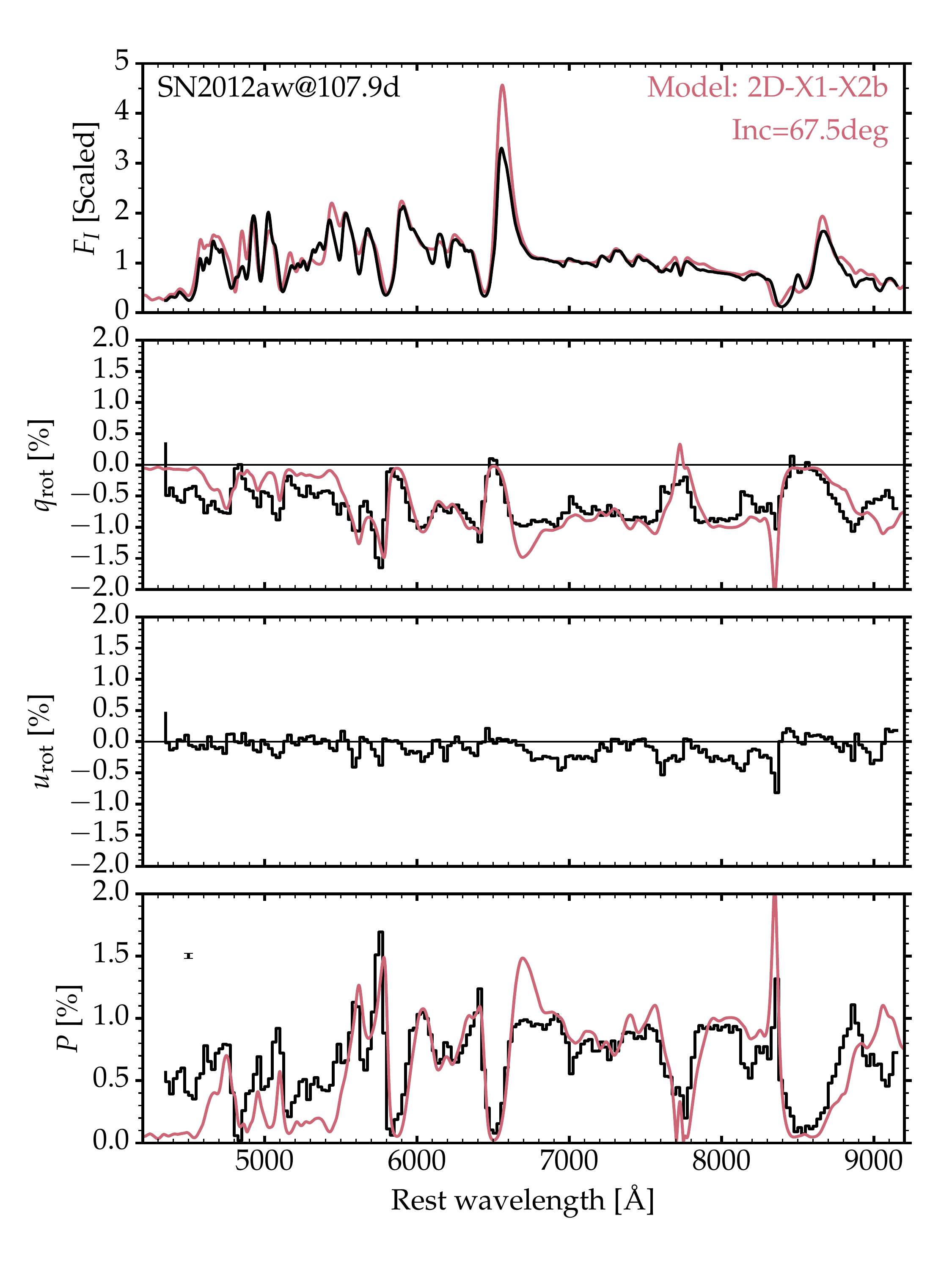, width=8.cm}
\epsfig{file=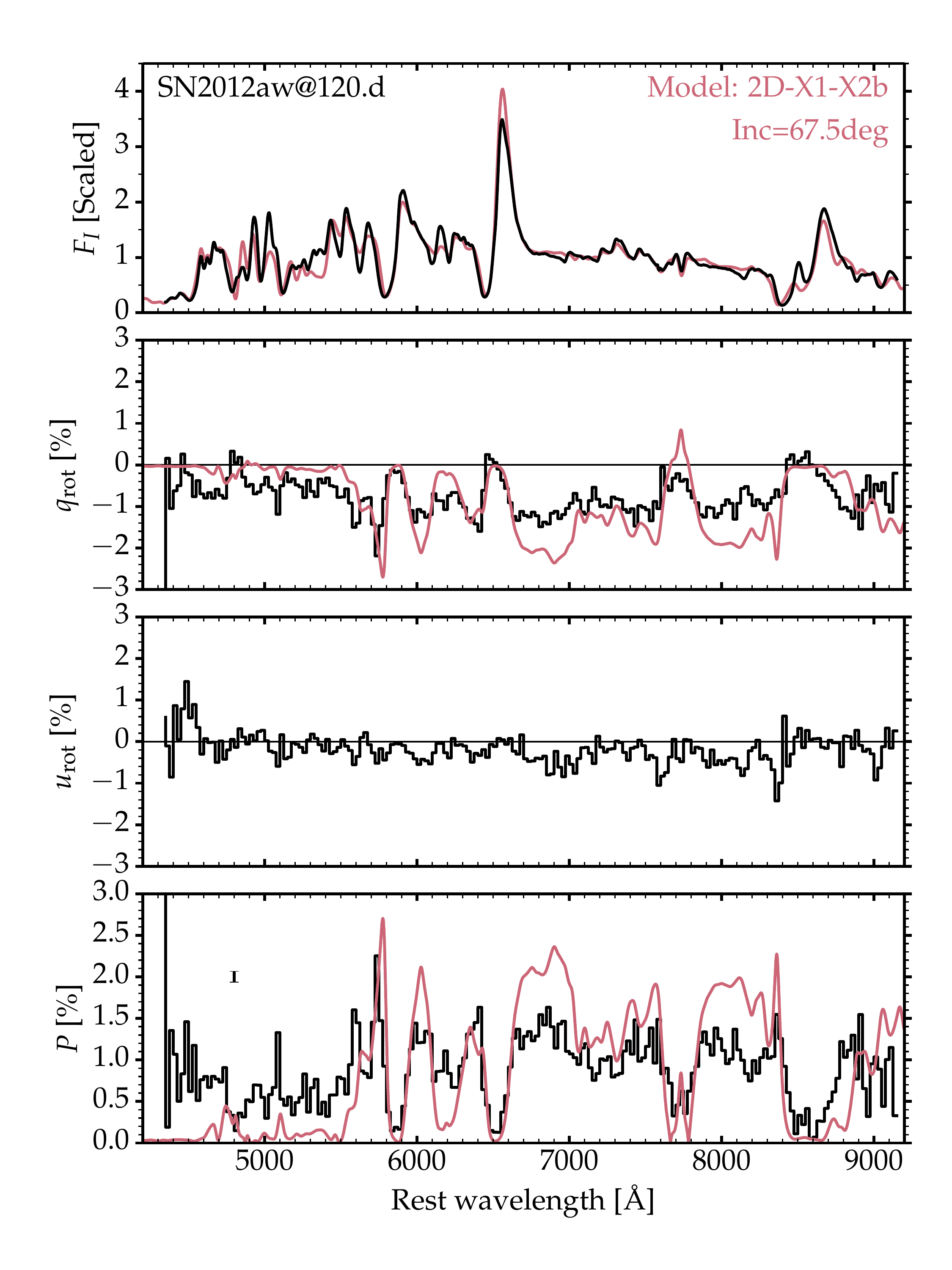, width=8.cm}
\epsfig{file=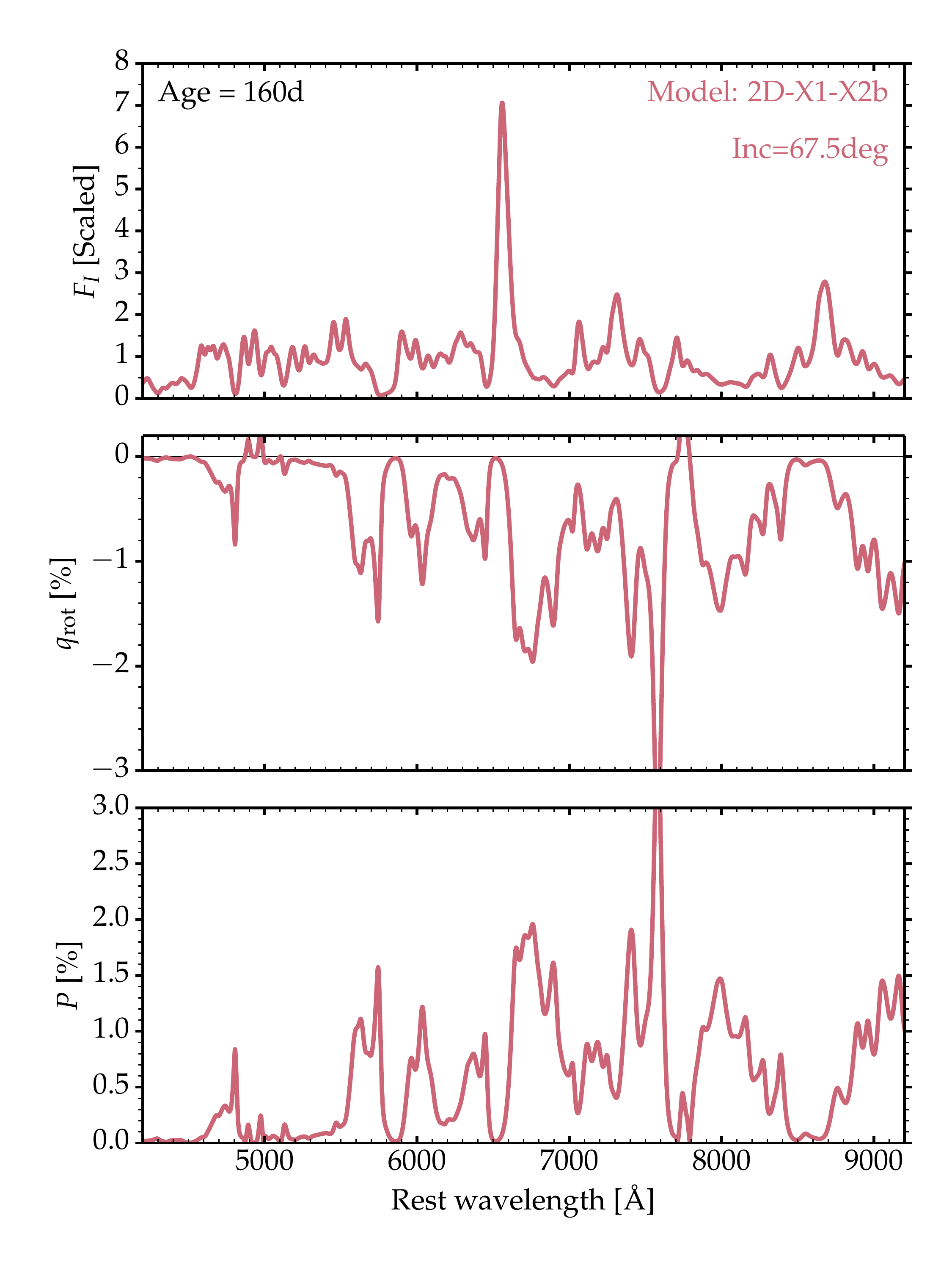, width=8.cm}
\end{center}
\vspace{-0.5cm}
\caption{Same as Fig.~\ref{fig_pol_1}, but now for epochs 5, 6, and 7. For the epoch 8 at 160\,d, there is no observational data for SN\,2012aw.
\label{fig_pol_2}
}
\end{figure*}

\subsection{Results with a bipolar explosion using a combination of 1D \cmfgen\ models}
\label{sect_multi}

In this section, we use two distinct 1D \cmfgen\ models to build an asymmetric ejecta. The ejecta are symmetric with respect to the equatorial plane so it corresponds to a bipolar explosion in which a cone of opening angle 50\,deg along the symmetry axis is characterized by distinct properties relative to lower latitudes. In our setup, the difference is the steeper density fall-off beyond about 10,000\,\kms\ and the larger \nifs\ content within the H-rich layers of the ejecta (see Fig.~\ref{fig_1d_prop}). In the context of polarization, the main impact of the \nifs\ enhancement is not the change in opacity or abundance (both have a minor influence) but instead the boost it induces to the density of free electrons (the mass density below about 10,000\,\kms\ hardly changes with angle, so the impact is on the ionization). Hence, there is at least one source of asymmetry at any given velocity, which implies that this setup may produce a net polarization at any epoch. The initial conditions used in this case are presented in detail in Section~\ref{sect_prep_lat_multi}, to which we refer the reader for additional information (see also Fig.~\ref{fig_1d_prop}\,$-$\ref{fig_init_2d_x1p5b3}).

One way to characterize the asymmetry of our ejecta is by comparing the photospheric radius along the pole and along the equator. Although our hybrid model is made of the same combination of 1D ejecta at all epochs computed, the gas properties evolve and in a distinct way along the polar and the equatorial directions. The \nifs\ enhancement affects mostly the innermost ejecta layers, so the asymmetry should be weak early on and grow as the inner aspherical ejecta layers are revealed by the receding photosphere. Consequently, in our hybrid model, the ratio of polar and equatorial photospheric radii evolves from 1.02, to 1.08, 1.22, 1.36, 1.59, 1.52 and 1.26 as the SN ages from 16.1, to 46.0, 63.0, 71.9, 91.9, 107.9, and 120.0\,d.

Another method for illustrating the structure is with the shape factor defined by \citet{Brown_McLean_77} which (together with the optical depth and the inclination) fully describes the continuum polarization properties of an axially-symmetric ejecta illuminated by a point source. In this paper, we introduce a depth-dependent shape factor  $\gamma(r)$ and define it as
\begin{equation}
\gamma(r) = { \int_r^\infty \int_{-1}^{1} N_e(r,\mu) \mu^2 \,d\mu \,dr \over \int_r^\infty \int_{-1}^{1} N_e(r,\mu)  \,d\mu \,dr\,}\,\,\, , \label{eq_gamma_r}
\end{equation}
where $N_e(r,\mu)$ is the free-electron density at $(r,\mu)$, and $\mu$ is the cosine of the polar angle $\beta$). When $r=R_{\rm min}$ we obtain the expression of the shape factor as in \citet{Brown_McLean_77}. Because of the complicated dependence of the emissivities and opacities on depth, no single parameter can fully describe the asymmetric nature of the ejecta. The shape factor (defined as in Eq.~\ref{eq_gamma_r}) for our hybrid model is shown for all epochs in Fig.~\ref{fig_gamma_r}.

Thus revealed, the level of asymmetry of the free-electron density distribution is modest at early times, but continuously grows as time progresses. This arises because the ionization level is naturally higher early on, irrespective of decay heating.\footnote{This is also why the shape factor is close to one third below the photosphere. In those regions of Type II SN ejecta, the conditions are ionized and electron scattering is the main opacity source.} However, when the ejecta recombines, the excess decay heating around 4000\,\kms\ and within the opening angle $\beta_{1/2}$ generates a local enhancement in free-electron density. The contrast grows in this region at all epochs and should boost the polarization level as time passes. Another striking property is that the asymmetry, as revealed by the shape factor, is really confined within the regions beyond 2000\,\kms, which are H rich. In contrast, the metal-rich core is moderately asymmetric.

In Figs.~\ref{fig_pol_1}\,$-$\,\ref{fig_pol_2}, we compare the multiepoch spectropolarimetric observations of SN\,2012aw with the results of the 2D polarized radiative transfer code based on this new axisymmetric ejecta configuration (the figure layout is the same as that used in Fig.~\ref{fig_pol_scale}). Figure~\ref{fig_pol_1} shows the results for the first four epochs (at 16.1, 46.0, 63.0, and 71.9\,d after explosion), and Fig.~\ref{fig_pol_2} shows the results for the last three epochs (91.9, 107.9, and 120.0\,d after explosion) together with the predictions of the model at 160\,d (for which there is no data). The later epochs are better suited for analyzing the SN polarization since the polarization level is higher and the data has a higher signal to noise ratio. The variations across lines are better resolved. For this comparison, we selected the inclination of 67.5\,deg since it yielded a good match to the observations at all epochs except the first one.

\begin{figure*}
\begin{center}
\epsfig{file=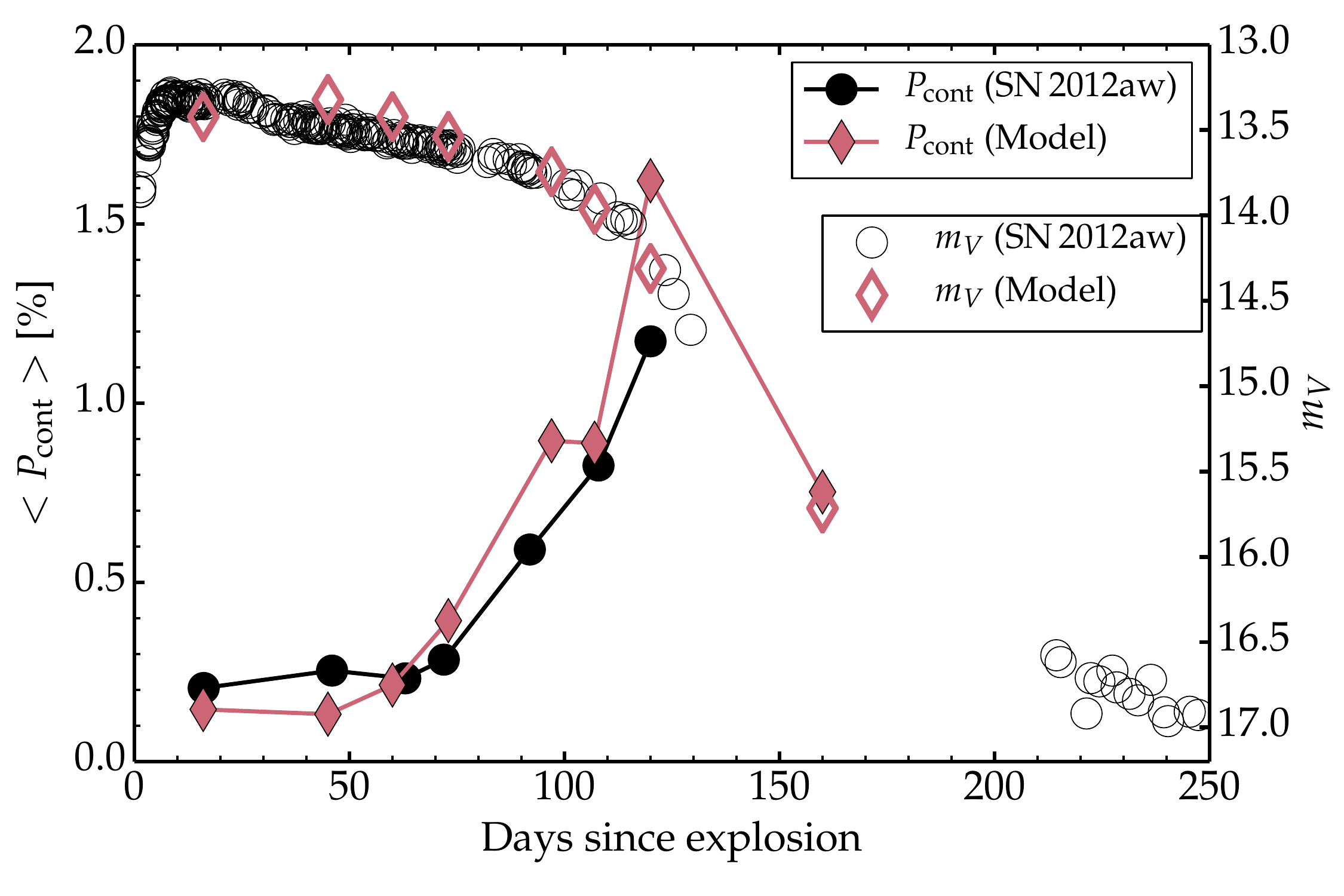, width=15cm}
\end{center}
\vspace{-0.5cm}
\caption{Evolution of the observed continuum polarization of SN\,2012aw, corrected for interstellar polarization (filled symbols; averaged value over the observed range 6900 to 7200\,\AA) and its $V$-band magnitude (open circles; values shown on right axis, from \citealt{bose_12aw_13}), along with model predictions (closed diamonds for polarization, open diamonds for photometry); the model photometry accounts for extinction and distance dilution as in \citet{HD19}. The symbols at 160\,d are for the model only since the last observation is at 120\,d. In this hybrid 2D axisymmetric ejecta model, the ratio of polar and equatorial photospheric radii evolves from 1.02, to 1.08, 1.22, 1.36, 1.59, 1.52 and 1.26 as the SN ages from 16.1, to 46.0, 63.0, 71.9, 91.9, 107.9, and 120.0 d. As discussed in Section~\ref{sect_isp}, a systematic uncertainty of $\sim$\,0.2\,\% exists on all continuum polarization measurements of SN\,2012aw displayed in this plot, due to the uncertainty in the ISP.
\label{fig_mv_pol_cont}
}
\end{figure*}

At the first two epochs, the observed polarization is low, about $0.1-0.3$\,\%. This polarization level is roughly matched by the hybrid model but the rise of the polarization at longer wavelength (more obvious at the first epoch) is not predicted (as for the previous calculations described in Sections~\ref{sect_scl} and \ref{sect_str}).  This rise in the red toward the Paschen jump is hard to comprehend as there are no obvious discontinuities in this region. If anything, the albedo decreases smoothly so the polarization is expected to drop as scattering weakens toward the red.  This peculiar feature may arise from a cancellation effect due to the competing influence of flux and optical depth (i.e., between the asymmetry of the emission and of the scatterers as seen on the plane of the sky).

At later epochs (63 to 107.9\,d), the hybrid model captures the qualitative and the quantitative evolution observed in SN\,2012aw, both for the total flux (and thus photometry) and for the polarized flux. This is obtained without any adjustment to the ejecta structure so that the change in polarization arises from the intrinsic evolution of the 2D ejecta as they expand and cool etc.

Across the profile of a strong line, the polarization is expected to show a jump in the blue part of the P-Cygni profile trough. This is observed in Na\one\,D and the Ca\two\ near-infrared triplet, but rarely in other lines. In contrast, the model shows this as a narrow feature in most lines, including H$\alpha$. The difficulty of observing this feature is aggravated by the rebinning of the data, which was used to enhance the signal-to-noise ratio per bin. The feature originates from the deficit of unscattered (direct and thus unpolarized) flux from the SN while the residual flux arises from photons scattered back into the line of sight.  In an asymmetric ejecta, such scattered photons yield a net polarization. In the Ca\two\ near-infrared triplet, the model shows the polarization jump further to the blue than in the observations. Close inspection of the total flux $F_I$ shows that the model overestimates the width of the Ca\two\ near-infrared triplet line at late epochs (more so at epochs 5, 6 and 7), so the mismatch in $q$ or $P$ is probably rooted in the slight overestimate by the model of the ejecta expansion rate.

In strong lines, the polarization is zero or close to zero somewhere between the P-Cygni trough and the location of maximum flux. This feature of line depolarization was used earlier to constrain the interstellar polarization toward the line of sight to SN\,2012aw. It is clearly observed at all epochs (more so for epochs three to seven) in H$\beta$, Na\one\,D, H$\alpha$, and the Ca\two\ near-infrared triplet. It is also reproduced by the model. This results from the depolarization of line photons. Recombination lines emit photons isotropically and thus cannot produce a net polarization. The large line optical depth also leads to absorption of possibly polarized photons, thereby destroying the polarization they carried. The line photons that scatter with free electrons within the ejecta can however produce a residual polarization. However, this is associated with a redshift (because of the large bulk velocity of the scatterers) so this polarized flux appears in the red wing of each line.\footnote{If the scattering was done by an external scatterer at rest, such as a distant dust sheet, line photons could be associated with a residual polarization even at line center since the scattering process would introduce no velocity shift.}

As we move to the red wing of lines, the polarized flux is large, and larger for stronger lines. This arises from line photons scattered by free electrons. The effect yields an excess total flux in the red wing of all lines. Because of the asymmetry of the distribution of free electrons, it also yields a clear polarized flux $F_Q$, whose strength scales with $F_I$. However, $P$ is at the same level in the red wing of lines and in the adjacent continuum.  This suggests that line photons and continuum photons pick up the same level of polarization when they last scatter in the ejecta (and that statistically they bear the same level of polarization before that last scattering). The observed polarization behavior across lines is quite distinct in type Ia SNe (the metal-rich composition implies a lower number of free electrons per unit mass and a reduced importance of electron scattering for the opacity), which show a maximum polarization within the absorption trough of the strongest lines and a low level of polarization elsewhere (see, e.g., \citealt{wang_01el_03}).

In general there is good agreement between the observed and theoretical line profiles, especially after the first two epochs. One discrepancy is that the model profiles exhibit a larger variation in polarization at the blue edge of the P~Cygni absorption than do the observations. Particularly striking is the agreement for the polarization across the O\one\,7774\,\AA\ multiplet. In the flux spectrum, O\one\,7774\,\AA\ is an inconspicuous spectral feature. However, in both the observed and model polarization (after the first two epochs) it is conspicuous, and is only slightly less prominent than the feature due to H$\alpha$.

In the continuum, the level of polarization increases with wavelength in the optical range, although this effect, which is matched by the model, is quite weak. At late times (epochs four to seven), in the region of line blanketing shortward of 5000\,\AA, the model predicts no clean (line free) continuum region so strong line depolarization occurs. In the red part of the spectrum, and in particular between H$\alpha$ and O\one\,7774\,\AA, the continuum polarization shows a smooth and near-constant magnitude both in the observations and in the model (the model shows fluctuations associated with weak lines, such as the Ca\two\,7300\,\AA\ doublet). This is the spectral region where it is maximum at any given epoch.

 Other discrepancies convey some important information on the source of polarization and departures from what is adopted in the hybrid model. Clearly visible at epochs  5 and 6 is the overestimate of the polarized flux associated with the scattered and red-shifted H$\alpha$ photons, appearing as a bump in $q$ or $P$. In contrast, the observations suggest the same level of polarization for these photons and the adjacent continuum-only photons further to the red. This mismatch is, however, only so obvious in H$\alpha$ and not, for example, in Na\one\,D. The origin of the discrepancy might be an inadequate \nifs\ distribution which causes a too extended boost to the electron density (for example, a more localized enhancement in \nifs\ at large velocity may be more suitable).

Another important discrepancy, which could be related to the point discussed in the previous paragraph, is the lack of model polarization in the blue part (say $<$\,5500\,\AA) of the spectrum at the recombination epoch (epochs 3 and later). The observations reveal a lower polarization than around 7000 or around 8000\,\AA, but only by about half. The model shows low or zero polarization because the polarized photons from the inner asymmetric ejecta are destroyed by interactions with lines (these regions are strongly affected by line blanketing due to Fe\two\ and Ti\two). One simple way of resolving this issue is to invoke a more distant, that is a more external scattering source, for example in the form a higher velocity \nifs\ enhancement. In this context, this external scatterer would yield a polarization that would scale with the number of incoming photons. The polarization in the blue could then be slightly smaller than at long wavelength because of the residual influence of line opacity at large velocities in the ejecta.

In the comparisons shown in Figs.~\ref{fig_pol_1} and \ref{fig_pol_2}, the observed polarization (second and third panels from top) was rotated so that the bulk of the polarization lies in the flux $F_Q$. Doing this, the flux $F_U$ is close to zero and shows only modest variations with wavelength, in contrast to what is observed in $F_Q$ (this is more easily seen at late epochs when the polarization in line-free regions reaches 0.5 to 1.0\%). At all epochs except the first one, the flux $F_Q$ is negative at all wavelengths in both the model and the observations. Together with the fact that $F_U$ is close to zero and much smaller (in magnitude)  than $F_Q$, this property suggests that the ejecta is primarily axisymmetric. This property is also evident from the data distribution in the $(q,u)$ plane shown in Fig.~\ref{fig_obs_q_u}.

Figure~\ref{fig_mv_pol_cont} summarizes the evolution of the continuum polarization (taken from the averaged value over the range 6900 to 7200\,\AA) as well as the $V$-band brightness for both SN\,2012aw and for the hybrid model discussed in this section. The model matches well the light curve for the plateau duration and brightness, and for the tail brightness. This occurs because the asymmetry has little effect on the total observed flux -- the hybrid model displays the same light curve and spectral evolution as the spherically symmetric model 1D-X1 (with the exception of a slight overestimate of the H$\alpha$ line strength and width). The hybrid model also matches the low level of polarization at early times and its steep rise in the second half of the plateau.  The hybrid model then predicts that the continuum polarization will drop at late times, essentially following the drop in optical depth. Under optically-thin conditions, the polarization scales linearly with $\tau$ \citep{Brown_McLean_77} and in SN $\tau$ drops as $1/t^2$ at late times (assuming constant ionization). Such an evolution has been observed over a sizeable timeline \citep{leonard_04dj_06} and explained by polarization modeling \citep{DH11_pol}.

\section{Uniqueness of the solution}
\label{sect_degen}

In the preceding section, we have explored various configurations for the aspherical (2D axisymmetric) ejecta. Here, we study the sensitivity of our results to changes in the 2D ejecta structure in the hybrid model. The hybrid model used model 1D-X2b for the polar latitudes (corresponding to a cone with an opening angle of $50-60$\,deg), and model 1D-X1 for other latitudes.  In this section, we describe the results when the half opening angle of the cone is increased (Section~\ref{sect_cone}), when we drop the assumption of mirror symmetry (equivalent to comparing a bipolar and a unipolar explosion; Section~\ref{sect_mirror}), and when the explosion energy of the ejecta in the polar regions is increased (Section~\ref{sect_expl}).

\subsection{Variation with cone opening angle in the hybrid model}
 \label{sect_cone}

\begin{figure}
\begin{center}
\epsfig{file=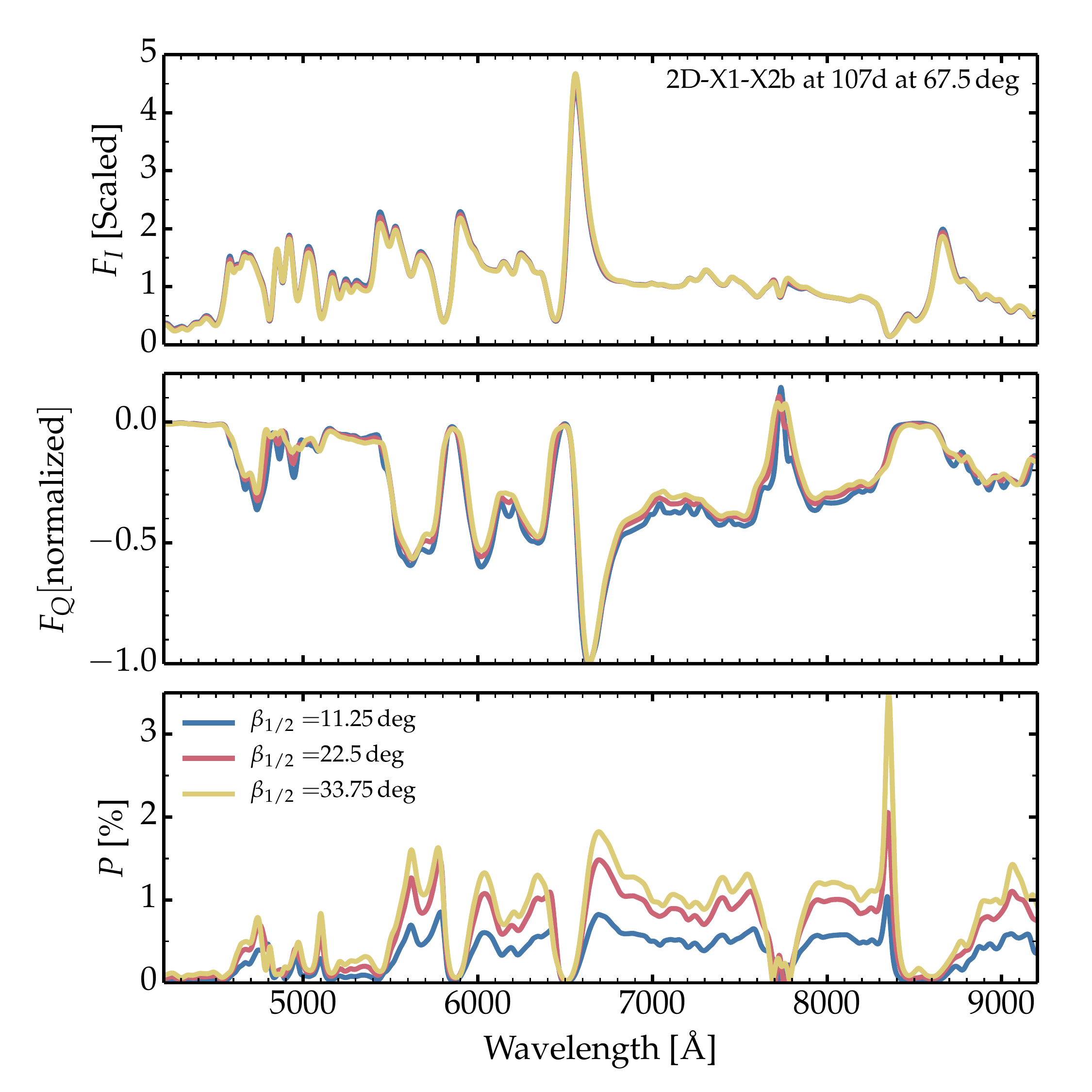, width=9cm}
\end{center}
\vspace{-0.5cm}
\caption{Influence of the half opening angle $\beta_{1/2}$ on the total flux $F_I$ (top), the normalized polarized flux $F_Q$ (middle) and the percentage polarization $P$ (bottom) for the hybrid model 2D-X1-X2b at 107\,d and using an inclination of 67.5\,deg. [See Section~\ref{sect_cone} for discussion.]
\label{fig_cone}
}
\end{figure}

 Figure~\ref{fig_cone} shows the influence of the half opening angle $\beta_{1/2}$ on the radiative signatures for model 2D-X1-X2b at 107\,d. In this particular setup, model X2b extends from the pole up to $\beta_{1/2}$ of 11.25, 22.5, or 33.75\,deg. For an inclination of 67.5\,deg,  we find that the normalized polarized flux (middle panel of Fig.~\ref{fig_cone}) is qualitatively independent of $\beta_{1/2}$, while quantitatively, the percentage polarization increases with $\beta_{1/2}$ (bottom panel of Fig.~\ref{fig_cone}). For lower inclinations, optical depth effects cause a sign flip (rotation by 90\,deg) of the polarization $F_Q$ for different values of $\beta_{1/2}$. Such optical depth effects complicate the analysis of the observed polarization during the photospheric phase. Furthermore, the polarization increase with increasing $\beta_{1/2}$ occurs up to a half opening angle of about 60\,deg, beyond which the configuration is somewhat equivalent to swapping model X1 and X2b for a half opening angle of $\pi/2 - \beta_{1/2}$.

\subsection{Unipolar versus bipolar explosion in the hybrid model}
 \label{sect_mirror}

\begin{figure}
\begin{center}
\epsfig{file=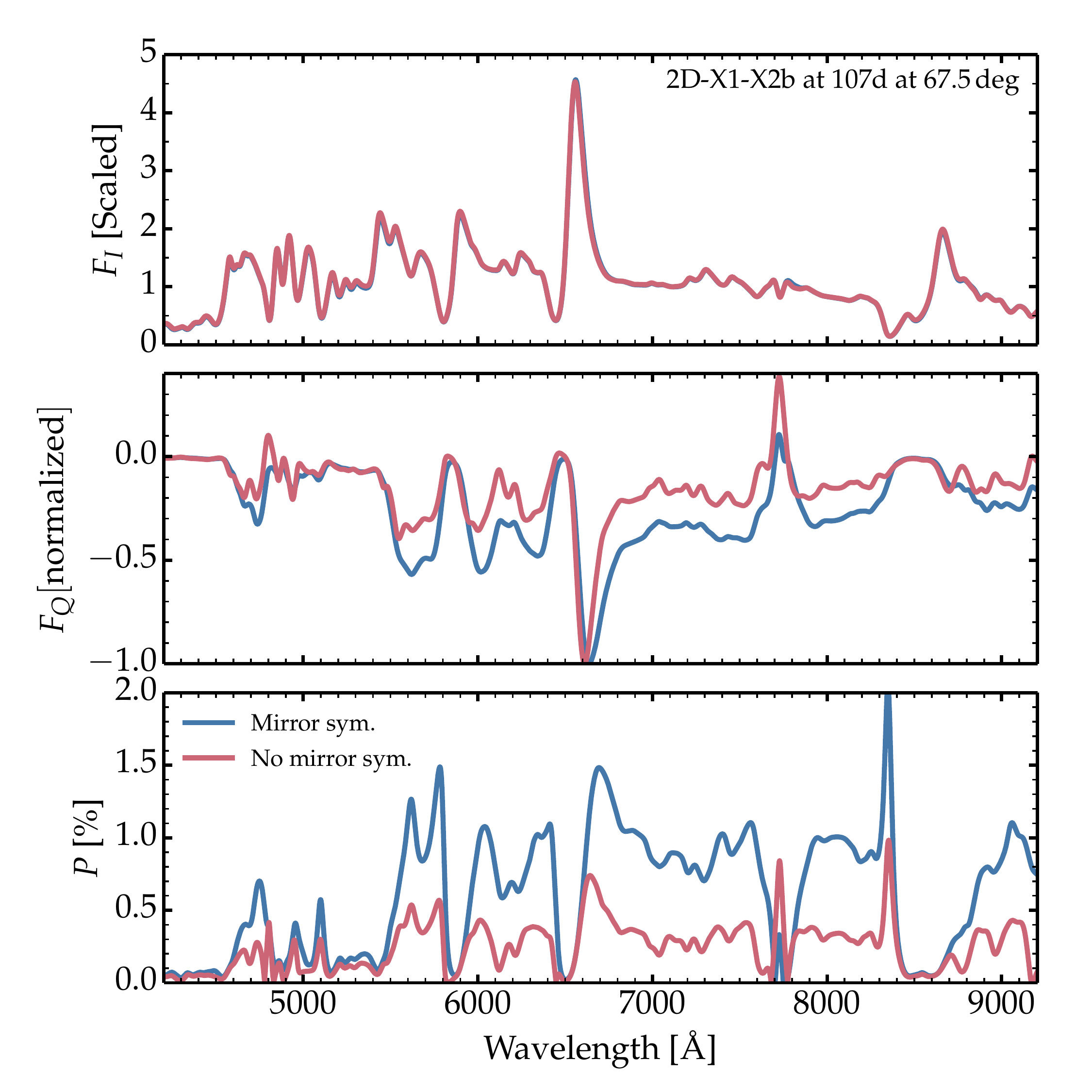, width=9cm}
\end{center}
\vspace{-0.5cm}
\caption{Same as Fig.~\ref{fig_cone}, but now showing the impact of the assumption of mirror symmetry. With (without) mirror symmetry, the ejecta (or the explosion) is bipolar (unipolar). [See Section~\ref{sect_mirror} for discussion.] \label{fig_mirror} }
\end{figure}

Figure~\ref{fig_mirror} shows the results for the hybrid model 2D-X1-X2b with and without the assumption of mirror symmetry. The angle $\beta_{1/2}$ is 22.5\,deg and the inclination is 67.5\,deg with respect to the symmetry axis. With mirror symmetry, the ejecta has a bipolar morphology, with the properties of model X2b along the poles. Without mirror symmetry, the ejecta is unipolar so that model X2b only covers polar angles smaller than 22.5\,deg. In Fig.~\ref{fig_mirror}, we adopt an inclination of 67.5\,deg. In this case, the model with mirror symmetry (i.e., bipolar explosion) shows a percentage polarization in the 7000\,\AA\ region that is more than twice greater than when mirror symmetry is ignored (i.e., unipolar explosion). The offset differs from a factor of two because of optical depth effects. For example, for an inclination of 10\,deg, the percentage polarization is the same in both cases because the asymmetry from the receding part of the 2D ejecta is obscured. For an inclination of 45\,deg, the level of polarization is very different (sign flip, different absolute polarization level) between the two cases, again because of optical depth effects. For an inclination of 90\,deg (edge on), the percentage polarization in the 7000\,\AA\ region is exactly twice larger if mirror symmetry is assumed rather than ignored. In the mirror-symmetry case, each hemisphere contributes the same residual polarization and the same flux at any wavelength. Without mirror symmetry, only one hemisphere contributes, yielding exactly half the continuum polarization level obtained with mirror symmetry. Finally, whatever the inclination, the total flux $F_I$ is essentially identical in both cases.

\subsection{Influence of explosion energy in the hybrid model}
\label{sect_expl}

\begin{figure*}
\begin{center}
\epsfig{file=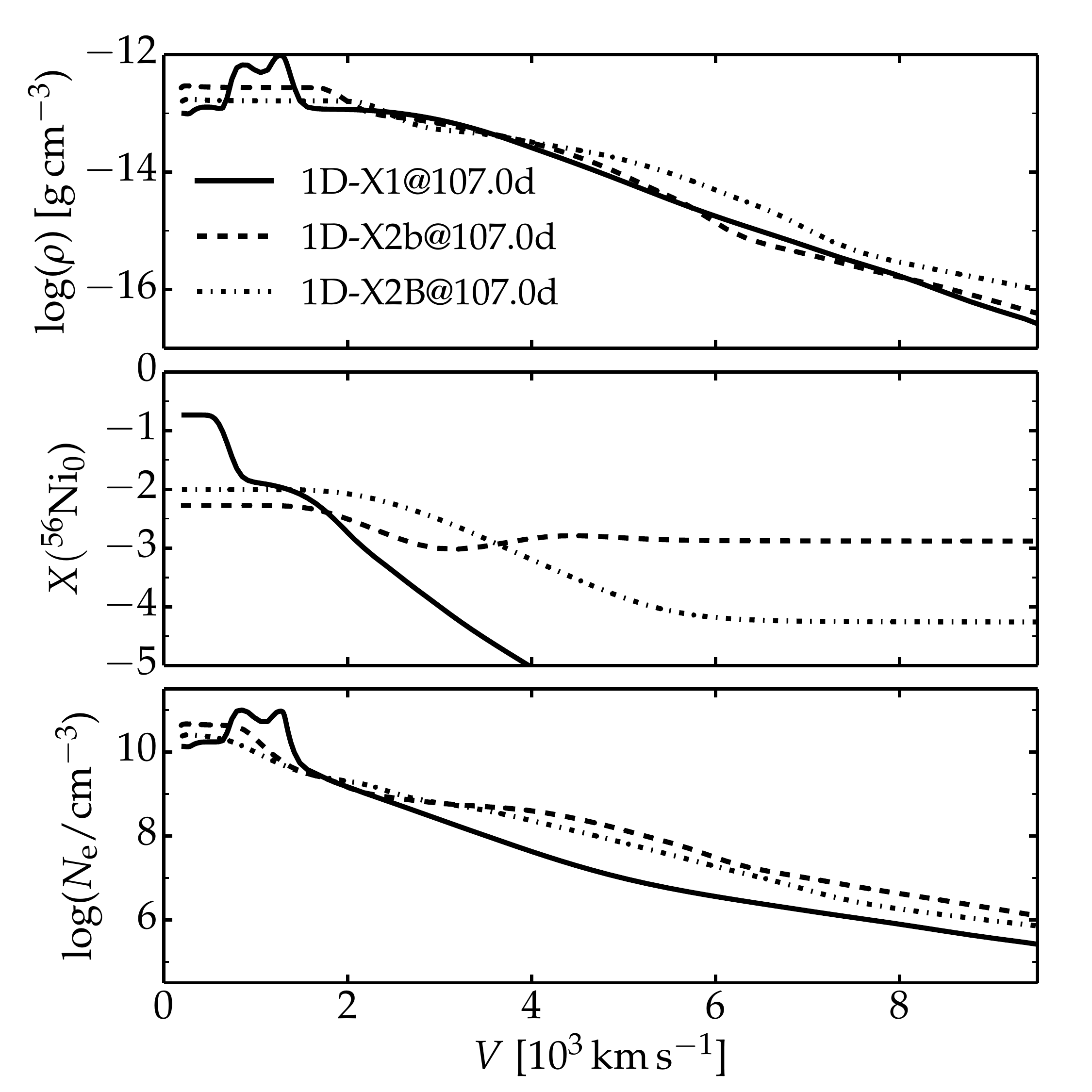, width=9cm}
\epsfig{file=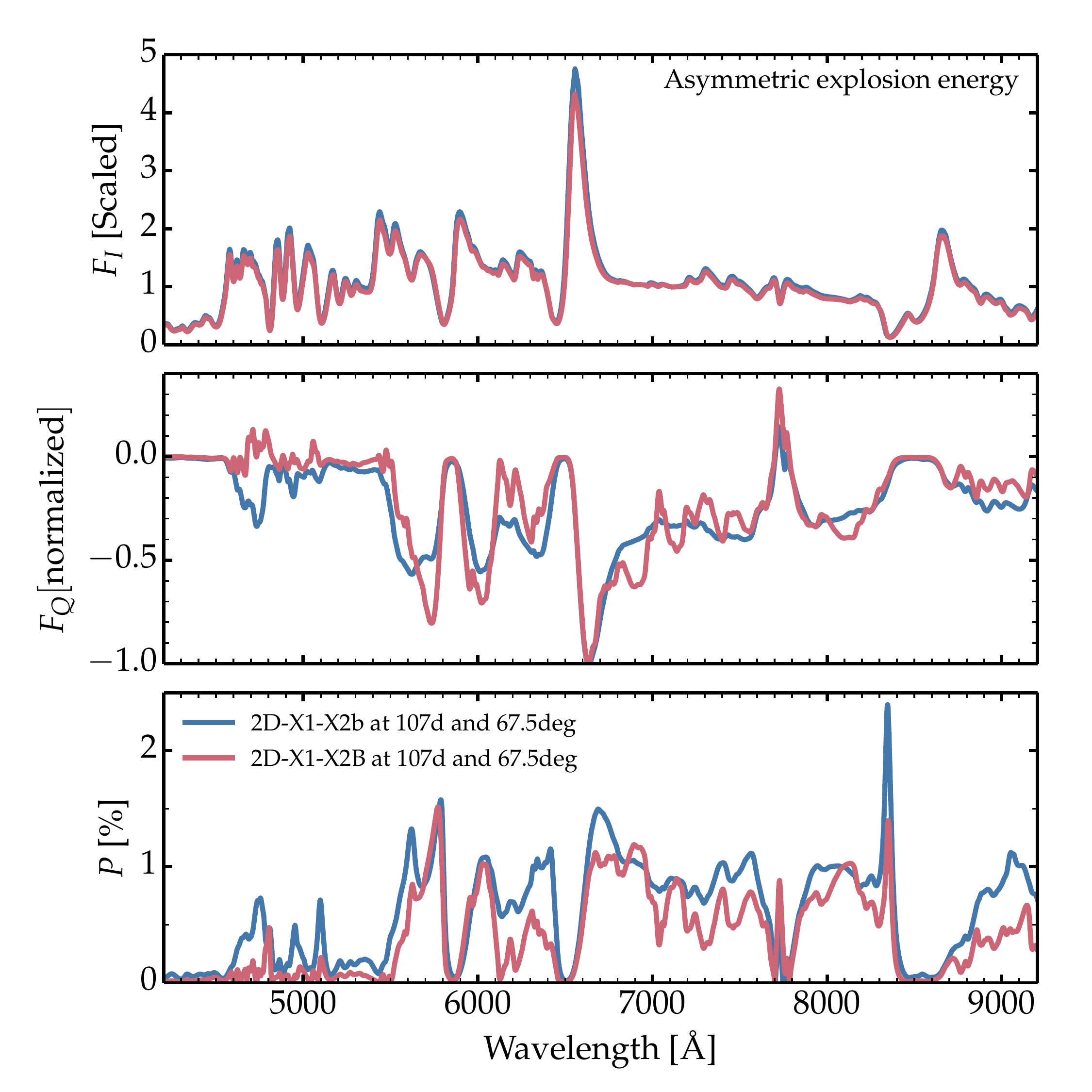, width=9cm}
\end{center}
\vspace{-0.5cm}
\caption{Left: Comparison of the mass density, \nifs\ mass fraction, and electron density for the 1D models used to build the 2D ejecta 2D-X1-X2b and 2D-X1-X2B. Right: Same as Fig.~\ref{fig_cone}, but now comparing the results for the hybrid model 2D-X1-X2b (bipolar ejecta with an enhanced \nifs\ mass fraction along the polar regions) and the hybrid model 2D-X1-X2B (bipolar ejecta with an enhanced explosion energy along the polar regions). In both models the angle $\beta_{1/2}$ is 22.5\,deg. [See Section~\ref{sect_expl} for discussion.] \label{fig_expl} }
\end{figure*}

Figure~\ref{fig_expl} compares the ejecta and radiative properties of the hybrid models 2D-X1-X2b and 2D-X1-X2B, which differ in that the polar regions have an enhanced \nifs\ mass fraction (former model) or an enhanced explosion energy (latter model). Although these two asymmetric ejecta are produced by very different means, they yield very similar total and polarized fluxes. The reason is that in the polar direction, the higher explosion energy leads to a greater density at large velocities. Although the ionization is the same along all latitudes (because of the similar influence of the deeply embedded \nifs), the electron density varies with latitude, reflecting the variation in mass density (see left panel of Fig.~\ref{fig_expl}). This polar enhancement in electron density (not caused by an excess in \nifs), produces the continuum polarization in model 2D-X1-X2B. In contrast, the mass density in model 2D-X1-X2b is essentially constant with latitude but the greater \nifs\ mass fraction along the pole yields a greater ionization and thus a greater density of free electrons along the poles than at lower latitudes. The results in Fig.~\ref{fig_expl} indicate that such distinct physical processes (polar enhanced \nifs\ mixing or polar enhanced explosion energy) yield a very similar polarization signal. This emphasizes the degeneracy of polarization signatures.

\section{Conclusion}
\label{sect_conc}

We have presented VLT\,$-$\,FORS spectropolarimetric observations of the type II SN\,2012aw spanning seven epochs of the photospheric phase until the onset of the drop-off from the plateau. Unfortunately, no spectropolarimetric observation of SN\,2012aw was made during the nebular phase. SN\,2012aw presents polarization characteristics that are qualitatively and quantitatively similar to what has been observed in type II SNe so far, such as the emblematic SN\,2004dj \citep{leonard_04dj_06}. A generic property of Type II SNe, corroborated here with 2012aw, is the relatively low polarization early in the plateau phase (although see the counter-example of SN\,2013ej; \citealt{leonard_iauga_15}; \citealt{mauerhan_13ej_17}, or SN\,2017gmr; \citealt{nagao_17gmr_pol_19}) and the progressive rise to a maximum polarization of about 1\,\% as the ejecta turns optically thin at the end of the high-brightness phase. By rotation of the Stokes vectors, it is possible to place most of the SN\,2012aw polarization along a single, fixed polarization axis, which implies that the ejecta of SN\,2012aw has a dominant axis and that the polarization arises from ejecta with an oblate or prolate morphology.

We have modeled the SN\,2012aw linear polarization using a long characteristic code that assumes axial symmetry \citep{hillier_94, hillier_96, DH11_pol,HD20_pol}. The code provides the full optical total and polarized flux $F_Q$ (for symmetry and geometry reasons, $F_U$ is zero) at any epoch during the photospheric or nebular phase. At present, the favored operation mode for the code is to build a 2D axially-symmetric ejecta using two 1D \cmfgen\ models and assigning them a specific range of latitudes. Such a hybrid 2D model can thus mimic an explosion with a higher energy or a higher \nifs\ mass fraction along specified latitudes.

Our modeling results support the notion that the intrinsic polarization is negligible within strong lines somewhere between the location of maximum absorption and the location of maximum line flux. We use this property to correct for the interstellar polarization. With this choice, the intrinsic polarization of SN\,2012aw is small but non zero at the earliest epochs.

The parameter space for producing 2D hybrid models is extended since we may use any 1D \cmfgen\ from the large set of models that we have calculated for type II SNe. We also have freedom when assigning a given model and a given latitude. Having selected two models and a specific model-latitude assignment to produce a 2D ejecta, the same setup is used for all epochs. No further adjustment is made to the setup during a sequence of polarization calculations. In the present paper, we focused on one hybrid model composed of model 1D-X1 (also named x1p5 in \citealt{HD19}) and of model 1D-X2b, which deviates from model 1D-X1 by having enhanced \nifs\ mixing at large velocities. For an inclination of $\sim$\,70\,deg, we find that this hybrid model can reproduce the SN\,2012aw polarization at all epochs except the first two, without any adjustment. The model predicts the polarization is maximum in the line free region between H$\alpha$ and O\one\,7774\,\AA, and also reproduces the variation in polarization across line profiles. The model predicts negligible polarization in line blanketed regions, while the observations suggest a nonzero residual polarization. However, the wavelength and time variation of the polarization is reproduced by the model, reflecting the internal changes in the corresponding ejecta (expansion, recession of the photosphere, cooling, recombination, reduction in density etc). As is well known \citep[e.g.,][]{jeffery_87_pol_91}, we also find that the polarization signatures are degenerate so other configurations can also reproduce the observations.

\begin{acknowledgements}

D.C.L. acknowledges support from NSF grants AST-1009571, AST-1210311, and AST-2010001, under which part of this research was carried out. D.J.H. thanks NASA for partial support through the astrophysical theory grant 80NSSC20K0524. Support for G.P. is provided by the Ministry of Economy, Development, and Tourism's Millennium Science Initiative through grant IC120009, awarded to The Millennium Institute of Astrophysics, MAS. This work was granted access to the HPC resources of  CINES under the allocation  2018 -- A0050410554 and 2019 -- A0070410554 made by GENCI, France. This research has made use of NASA's Astrophysics Data System Bibliographic Services. Based on observations collected at the European Southern Observatory, Chile, under programme 089.D-0515(A).

\end{acknowledgements}

\end{document}